# NON-EQUILIBRIUM QUANTUM FIELDS IN THE LARGE N EXPANSION

Fred Cooper, Salman Habib,
Yuval Kluger, Emil Mottola, and Juan Pablo Paz

*Theoretical Division*
*Los Alamos National Laboratory*
*Los Alamos, New Mexico 87545*

and

Paul R. Anderson

*Department of Physics*
*Wake Forest University*
*Winston-Salem, North Carolina 27109*

## Abstract

An effective action technique for the time evolution of a closed system consisting of one or more mean fields interacting with their quantum fluctuations is presented. By marrying large $N$ expansion methods to the Schwinger-Keldysh closed time path (CTP) formulation of the quantum effective action, causality of the resulting equations of motion is ensured and a systematic, energy conserving and gauge invariant expansion about the quasi-classical mean field(s) in powers of $1/N$ developed. The general method is exposed in two specific examples, $O(N)$ symmetric scalar $\lambda\Phi^4$ theory and Quantum Electrodynamics (QED) with $N$ fermion fields. The $\lambda\Phi^4$ case is well suited to the numerical study of the real time dynamics of phase transitions characterized by a scalar order parameter. In QED the technique may be used to study the quantum non-equilibrium effects of pair creation in strong electric fields and the scattering and transport processes in a relativistic $e^+e^-$ plasma. A simple renormalization scheme that makes practical the numerical solution of the equations of motion of these and other field theories is described.

# 1 Introduction

The derivation of macroscopic dissipative behavior from fundamental time reversal invariant dynamics is a subject at least as old as Boltzmann. In the classical approach to this issue two ingredients are necessary. First, non-equilibrium initial conditions for the system are considered, and second, some kind of coarse-graining or averaging procedure over unobserved degrees of freedom is introduced [1]. The classical prototype for this analysis is of course, Boltzmann's equation for the evolution of the molecular distribution function in a gas due to collisions between the molecules. Another classic example is the Brownian motion of a heavy particle in a fluctuating medium, where a clean separation between the "system" and "environmental" degrees of freedom to be averaged is assumed in the first step. In both of these examples from classical statistical mechanics the phenomenological point of view is the dominant one, and there is no question about how entropy growth or loss of information arises. It arises either by explicitly replacing the underlying time reversal invariant dynamics by stochastic assumptions, or by the separation into "system" and "environment" with all detailed information about the latter lost in the averaging procedure (save for a few parameters, such as temperature).

In recent years there has been considerable growth in interest in the study of dissipative effects in specifically quantum mechanical systems. Driven largely by the impressive progress in the fabrication and control of sensitive microdevices, such as tunnel junctions, issues of quantum dissipation have been studied extensively by a variety of methods [2]. At the other end of the distance scale, the subject of loss of coherence in closed quantum systems has become of interest in quantum cosmology, where one would like to understand precisely how the quasi-classical universe we observe emerged from (presumably) purely quantal initial conditions [3]. Because of their technical simplicity and applicability to these widely different situations, model theories of quantum Brownian motion have often served as prototypes in studies of dissipation and decoherence. In fact, almost all existing treatments of non-equilibrium dissipative phenomena in quantum systems focus on some particular model or class of models, such as the coupled oscillators of Caldeira and Leggett [4], chosen primarily for the purpose of illustrating some definite feature of the analysis in the simplest way, or for describing a particular existing phenomenology. Since relativistic effects are unimportant in condensed matter applications, they have been generally neglected in those phenomenological models.

There is no essential problem in extending the analyses of non-equilibrium and irreversible processes to systems with both quantum and relativistic features, *i.e.*, to relativistic quantum field theories. Such an analysis is not necessary for the primary application of field theories to experimental situations, namely scattering experiments involving a few particles, and this explains why it has not been much discussed in the literature. However, implicitly assumed in the calculation of a scattering matrix or cross section is that the particles in both the distant past (the *in* state) and in the remote future (the *out* state) of the scattering event are to be treated as essentially *free* particles. It is only this quite restrictive assumption that justifies the usual Feynman boundary conditions on Green's functions, and the corresponding path integral representation of the scattering matrix between *in* and *out* states.

As successful as this scattering theory has been, it is clear that it is capable of addressing only a very limited subclass of questions one might hope to address in quantum field theory. The coupling may not be weak enough to justify a perturbative expansion, there may be no approximately free asymptotic scattering states, or we may be interested in more than just scattering probabilities, and of course, such a treatment based on asymptotic states cannot teach us anything directly about dissipation or loss of quantum coherence. Only relatively recently has it become apparent that there are many situations of physical interest requiring a detailed description of the *time-dependent* dynamical evolution, where the perturbative scattering formalism of quantum field theory is wholly inadequate. For example, this is the case for strong field electrodynamics in astrophysical plasmas, in nuclear collisions of heavy ions where it is possible that a phase transition to the quark-gluon plasma may be produced, in phase transitions in the hot, dense early universe, and in the dynamics of phase transitions generally. All of these problems require a detailed knowledge of the dynamical time evolution of the field configurations, as well as their non-equilibrium transport and energy-momentum flow characteristics. It is at that point also that the fundamental issue of dissipation in the time reversal invariant physics described by the underlying quantum fields asserts itself.

Despite the apparent need for a time-dependent (and therefore not explicitly covariant) formulation of quantum field theory, to address both the fundamental issues and the many interesting applications, the technical complications of the renormalization procedure in a non-covariant treatment have usually been sufficient to dissuade all but the most persistent from the task. When renormalization issues are faced squarely, it is usually by rather formal methods, such as dimensional continuation. Although elegant theoretically this formal regularization technique is not adaptable to a computer, to which it should be clear from the outset one will have to resort to in order to solve the coupled non-linear partial differential equations of any realistic theory. In fact, it is only the advent of modern supercomputers that makes it even conceivable to carry out the program outlined in this paper and obtain useful results in realistic physical systems.

Motivated by these various considerations we wish in this paper to provide a fully quantum field theoretic framework for the systematic inclusion of the effects of fluctuations on the time evolution of a closed system in a formulation practical for numerical methods. In terms of our original discussion of "system" and "environment," the mean field (or fields) will play the role of the macroscopic subsystem, while the fluctuating degrees of freedom become the environment or bath in which the system moves. The important difference in our approach will be that the latter will not be introduced externally, but rather *determined* from the correlation functions of the theory self-consistently in terms of the time evolving mean field(s). In other words we wish to discuss the quantum evolution of a *closed* system, where *no* averaging over environmental degrees of freedom is ever performed. It is clear that the *exact* solution of this problem could never allow for any dissipation or decoherence, since well-defined Hamiltonian and unitary time evolution operators exist in the underlying field theory. However, we cannot hope to solve the Schwinger-Dyson equations of a non-trivial interacting quantum field theory exactly, so that some systematic approximation scheme is certainly required. In statistical mechanics the natural approximation scheme is to truncate the infinite hierarchy of correlation functions at some finite order. It is at this point when higher order correlations are neglected that information about the exact unitary evolution may be lost, and *effective* time irreversibilty may enter. This interesting point has been emphasized by Calzetta and Hu [5].

The problem of the time evolution of a closed system together with its self-generated quantum fluctuations is clearly very general, and arises in a number of contexts in very different areas of physics. Original motivations in microscopic and mesoscopic quantum devices have been mentioned [2], but there are many other possible applications. To consider but a few examples, in electromagnetic plasmas the classical phenomenon of *radiation reaction* has its quantum counterpart when particle creation and virtual processes are considered. These lead to the damping of plasmon modes in strong fields, such as those in the vicinity of rapidly rotating neutron stars [6]. In heavy ion collisions of ultrarelativistic nuclei a strongly interacting quark-gluon plasma phase may be produced [7]. The theoretical understanding of nonequilibrium effects in such a plasma is very rudimentary at present, but is clearly required to interpret the data soon to be generated at accelerators such as RHIC or the LHC. Similar effects are expected in the real time dynamics of any phase transition, whether occuring in terrestrial condensed matter systems or in the early universe. In black hole evaporation, the effect of fluctuating degrees of freedom on the mean field (the metric of spacetime) is called *backreaction*, and has given rise to much speculation on the status of entropy and information loss in quantum gravity. A fully consistent causal formulation of the time evolution problem for quantum field theory from given initial conditions is absolutely necessary to discuss this wide variety of quantum backreaction problems.

Apart from their backreaction on the evolution of the mean value, the fluctuating degrees of freedom are responsible for another important effect. In standard treatments the environment also induces a process of negative selection in the Hilbert space of the system: most of the quantum states become very unstable and rapidly decay into *mixtures* of relatively stable states. In other words, the environment induces classical behavior in the quantum system by dynamically suppressing the interference effects between macroscopically distinguishable states of the system. This is the *decoherence* process [8], which plays an essential role in the transition from quantum to classical behavior. As one example of this phenomenon, in a theory with spontaneous symmetry breaking, the coherence between the components of the wave function corresponding to different ground states decohere due to the coupling between the system and the environment, so that after some characteristic decoherence time the state is a classical mixture (and not a quantum superposition) of macroscopically distinguishable states corresponding to different vacua. As soon as the dynamics is forced to choose between macroscopically different values (as in spontaneous symmetry breaking potentials) then we have the essential ingredient for macroscopic quantum coherence. The decoherence time scale arising from interactions with the fluctuating field(s) may be studied in a such a situation. The importance of decoherence in the context of quantum field theory, the physics of the early universe and quantum cosmology has been recognized and studied in recent years [3]. Most of these studies focus on very simple quantum mechanical toy models and very little has been done in realistic quantum field theories or in closed systems generally. The techniques we develop in this paper are directly applicable to the study of such processes, since the universe as a whole is certainly a closed system.

The first requirement for the study of non-equilibrium time evolution is a general initial value formulation of quantum field theory. In several earlier papers, two of the authors have provided the necessary Heisenberg picture formulation of the initial value problem in the leading order of the large $N$ expansion [9]. It has been successfully applied to the problem of pair creation in strong electric fields [10]. This leading order approximation in $1/N$ is equivalent to the Hartree-Fock mean field approximation which has been much studied in nuclear many-body, atomic and molecular chemistry applications [11]. It corresponds to a Gaussian ansatz for the Schrödinger wave functional. In general relativity the leading order approximation consists of replacing the energy-momentum source on the right side of Einstein's equations by its expectation value, thereby ignoring the effects of fluctuations of $T_{\mu\nu}$ from its mean on the quasi-classical metric of spacetime. It has been clear for some time that this approximation is not adequate in the final stages of black hole evaporation from the Hawking effect, or in very early stages of the universe's expansion. However, technical difficulties have thwarted attempts to go beyond the Gaussian ansatz for wave functionals. Our main contribution in this paper is a systematic technique for doing just that, in a way suitable for practical numerical solution on a computer.

It is true that by suitably generalizing the Gaussian ansatz one should be able to evade the limitations of the mean field approximation in the Schrödinger picture as well. Such an approach is possible in principle but unattractive (in our view) principally because of the technical complications of the renormalization program and the loss of all connection to a covariant analysis of divergences. For example, in $\Phi^4$ theory, even in the lowest order approximation the Ward identities are not preserved by Gaussian trial wave functions without imposing the large $N$ limit as well. In general, variational ansätze at the level of wave functions do not necessarily preserve the invariances of the underlying theory. It is the issue of renormalization on which most previous forays into this area founder in a web of intricate technicalities.

In this paper we remove the restrictions of the Gaussian or mean field approximation by working in the Heisenberg picture, and making use of the Schwinger-Keldysh closed time path (CTP) formulation of quantum field theory [12]. It is this method that provides the technical means to formulate the initial value problem in a completely causal manner, removing the Feynman boundary conditions on Green's functions, which are only appropriate for $\langle out|in\rangle$ matrix elements and asymptotic scattering states. Otherwise, the techniques employed are completely familiar in field theory. Specifically, we make use of a (suitably modified) path integral representation for the generating function of connected Green's functions, and perform the Legendre transform to the effective action functional, which makes the covariance properties of the theory manifest. Hence, all conservation laws of the classical theory (which are not anomalous) are maintained explicitly. In particular, a conserved energy-momentum tensor for the fields in the plasma is obtained automatically by the effective action technique, and the renormalization program is no more difficult than in the ordinary covariant analysis. The variation of the effective action provides the dynamical evolution equation(s) for the quasi-classical mean field(s) and their fluctuations, in a form suitable for direct numerical integration from specified initial data at $t = 0$. Needless to say these equations of motion of the $\langle in|in\rangle$ expectation values of fields are real (which they would not be if the usual Feynman Green's functions were used).

The large $N$ expansion provides a convenient way of parameterizing the separation into quasi-classical mean fields and their fluctuations, and at the same time, a systematic approximation scheme to the Schwinger-Dyson equations of the full field theory. The method we espouse is therefore a kind of quantum analog of truncation of the BBGKY hierarchy of $n$-particle correlation functions of classical statistical mechanics at a finite order to render the system tractable to analysis [14]. Information loss and effective time irreversibility may enter because of this truncation [5], which is well motivated by the impracticality of having unlimited information about very high order correlations in the the initial state of any realistic system with many degrees of freedom. While the large $N$ method may not be the only approximation scheme that makes this separation possible, it is a gauge and renormalization group invariant expansion which permits non-perturbative mean fields [15]. In other words, the large $N$ method goes far beyond a simple perturbative expansion in the coupling constant(s) which is necessarily an expansion around the vacuum field configuration, such as is entirely inappropriate when energy densities are high and the state is far from the vacuum. The large $N$ expansion also resums and rearranges the Feynman perturbation series for scattering diagrams (collisions in the Boltzmann language) in a way that automatically includes self-energy corrections generated by the very same microphysical scattering and collisional processes. This points the way to a systematic quantal generalization of classical transport theory even in the presence of strong mean fields, where quantum self-energy (and all vacuum polarization or off-shell virtual processes) are treated on the same footing as collisional or real particle creation processes in the plasma. Fields which scale like $N$ to a positive power for large $N$ can be considered strong quasi-classical mean fields in this approach. Particle creation effects are contained already in the leading order (*i.e.*, $N^0$) approximation. In the next to leading order $(1/N)$ the effects of collisional, virtual and radiation reaction processes back on the quasi-classical fields appear for the first time. For almost all applications of physical interest, these effects are essential, and cannot be described by Gaussian wave functions. Their description requires the full power of the Schwinger-Keldysh initial value formulation of quantum field theory.

In order to expose the general method in as clear as possible a manner we shall consider in this paper two specific field theories for definiteness, scalar $\lambda\Phi^4$ theory and quantum electrodynamics (QED). Each of these is interesting in its own right. The $\lambda\Phi^4$ theory may be applied to the study of the time development of phase transitions characterized by a scalar order parameter, whether in the early universe or in a laboratory environment. In particular, the method is directly applicable to the chi-ral phase transition and the evolution of a disoriented chiral condensate in collisions of relativistic heavy nuclei [17]. Electrodynamics is interesting first because QED is arguably the most completely verified theory of fundamental interactions known and an excellent testing ground for general notions of dissipation or decoherence, and second, for application to particle production processes in strong field astrophysical plasmas. Although the details of the technique will be discussed in scalar $\Phi^4$ theory and QED in this paper, we would like to emphasize that the technique itself is ideally adapted to addressing the interesting fundamental issues of quantum dissipation, decoherence and time irreversibility in quantum field theories generally. The extension of these same methods to non-abelian gauge theories such as quantum chromodynamics (QCD), for application to the non-equilibrium time evolution of the quark-gluon plasma, and to gravity, for application to particle production processes in the very early universe and by black holes is planned in future publications.

We are encouraged to believe that our methods will be applicable to all these problems by an additional technical advance in the area of renormalization. In earlier papers on the initial value formulation of quantum field theory (ours included), an adiabatic WKB expansion in the time variation of the frequency of oscillation of the fluctuating quantum modes is employed [10][18]. The ultraviolet divergent contributions in the effective equations of motion appear in the first few orders of this asymptotic expansion and may be removed by explicit subtractions. In trying to extend this method beyond the lowest order mean field approximation and/or to spatially non-uniform mean fields we found that it quickly becomes very complicated, and ill-suited to practical calculations. Fortunately, we are able to dispense with this adiabatic expansion entirely, by the simple device of introducing an ultraviolet cutoff on formally divergent integrals (which one always does in practice on the computer in any case). Then we have only to check that rescaling the cutoff and flowing the coupling constant according to the standard continuum renormalization group to a given order in $1/N$ leaves physical, renormalization group invariant quantities unchanged. If the scale of non-equilibrium time evolution and other physics of interest is far from the cutoff scale this procedure is a very sensible one on physical grounds, and certainly it is much easier to verify that this procedure works *a posteriori* than it is to apply adiabatic type expansions beyond the lowest order in $1/N$. We give some explicit numerical evidence of the practical feasibility of this method in this paper.

In order to make the paper as self-contained and readable as possible, we review in the next two sections first the derivation of the effective action for the mean fields in the large $N$ expansion for both $\Phi^4$ theory and QED, and second, the basics of the Schwinger-Keldysh real time CTP formalism, from the functional integral point of view. In Section 4 we bring the two ingredients together and derive the causal equations of motion for both scalar $\Phi^4$ theory and electrodynamics in the large $N$ expansion. Section 5 is devoted to consideration of renormalization issues for the equations of motion, conserved currents, and energy-momentum tensors of each theory. We present here preliminary numerical evidence of the practicality of the renormalization scheme without the cumbersome adiabatic expansion employed in earlier papers. Our conclusions are summarized in Section 6.

## 2  The Large N Expansion

Scalar field theory with a $\lambda \Phi^4$ self interaction is the simplest renormalizable quantum field theory with which to develop the techniques for separating a closed system into mean fields and the fluctuations about them. This model is interesting in its own right for the study of phase transitions characterized by a scalar order parameter, and for its role in the Higgs sector of the standard model of electroweak interactions. Let us begin by reviewing the large $N$ expansion in the context of an $N$-component real scalar field with the $O(N)$ invariant Lagrangian density [15],

$$L = -\tfrac{1}{2}\Phi_i G^{-1}[\chi]\Phi_i + \frac{N}{\lambda}\chi\left(\frac{\chi}{2} - \mu^2\right), \tag{2.1}$$

where $i = 1,\ldots,N$ and

$$G^{-1}[\chi] \equiv -\Box + \chi . \tag{2.2}$$

This form is equivalent to the Lagrangian density,

$$L' = -\tfrac{1}{2}(\partial_a\Phi_i)(\partial^a\Phi_i) - \tfrac{1}{2}\mu^2(\Phi_i\Phi_i) - \frac{\lambda}{8N}(\Phi_i\Phi_i)^2 \tag{2.3}$$

with the definition of the composite field $\chi$ by

$$\chi = \mu^2 + \frac{\lambda}{2N}\Phi_i\Phi_i, \tag{2.4}$$

since the two Lagrangians $L$ and $L'$ differ only by a constant and a surface term. The quartic coupling in the Lagrangian has been taken to be $\lambda/N$ from the outset, rather than rescaling it later by $1/N$ as is sometimes done [16].

Adding independent sources to the Lagrangian and integrating over each of the fields $\chi$ and $\Phi_i$ defines the generating functional,

$$\begin{aligned}Z[J,K] &\equiv \exp(iNW[J,K]) \\ &\equiv \int[\mathcal{D}\chi]\prod_{i=1}^{N}\int[d\Phi_i]\exp\left\{i\int d^4x\ (L[\Phi,\chi] + J_i\Phi_i + NK\chi)\right\}.\end{aligned} \tag{2.5}$$

Here factors of $N$ have been inserted which multiply both the $\chi$ source $K$ and the generating functional for connected Green's functions $W$, in anticipation of the large $N$ expansion which we wish to examine. Because of the introduction of the composite field $\chi$, the integrations over the $N$ scalar fields $\Phi_i$ are Gaussian and may be performed explicitly:

$$\begin{aligned}
Z[J,K] &= \int [\mathcal{D}\chi] \exp\left\{i \int d^4x \left[\frac{N}{\lambda}\chi\left(\frac{\chi}{2} - \mu^2\right)\right]\right\} \\
&\quad \times \exp\left\{i \int d^4x \left[NK\chi + \frac{1}{2}\int d^4y\, J_i(x) G[\chi]_{ij}(x,y) J_j(y)\right]\right\} \\
&\quad \times \exp\left\{-\frac{N}{2}\operatorname{Tr}\ln G^{-1}[\chi]\right\},
\end{aligned} \quad (2.6)$$

where the notation $G[\chi]_{ij}(x,y) = \delta_{ij} G[\chi](x,y)$ for the inverse of $G^{-1}[\chi]$ has been used. If we take each of the $N$ copies of the original scalar field $\Phi$ to be equivalent, we may set each of the $N$ components of the source $J_i$ to be equal, and write

$$\int d^4x \int d^4y\, J_i(x) G[\chi]_{ij}(x,y) J_j(y) = N J \circ G \circ J \quad (2.7)$$

in a more compact notation. In all of the following, the symbol $\circ$ will denote summation over internal indices and integration over continuous spacetime coordinates in the quantities on either side of it (the de Witt summation convention), whereas omission of the $\circ$ will denote simple multiplication without summing or integrating over the coordinate labels.

Once the generating functional (2.6) has been obtained by integrating over the $N$ copies of the scalar field, (2.7) is obviously equivalent to simply rescaling the source $J$ for a *single* $\phi$ field (and the $\phi$ field itself) by $\sqrt{N}$. This is the sense in which the $\phi$ field is quasi-classical, since its mean value is strongly enhanced with respect to ordinary perturbative treatments of the quantum field. A large value of the classical field strength implies that we are expanding about a field configuration far from the perturbative vacuum, which is the reason that the large $N$ expansion is useful for matter under extreme conditions of high density or temperature. We should remark in passing as well, that scaling all $N$ copies of the scalar field in the same way is *not* appropriate if one is interested in spontaneous breaking of the $O(N)$ symmetry, in which case one component should be singled out and treated differently from the remaining $N-1$ Goldstone fields. We shall ignore this distinction in the following, and focus on the $O(N)$ symmetric case in order to simplify the presentation.

It should be clear now why we introduced the factors of $N$ as we did, for the exponent of the integrand in (2.6) contains an explicit overall multiplicative factor of $N$, and the $\chi$ integration may be performed by the stationary phase method in the limit of large $N$. The stationary phase point of the integrand $\chi_s[J,K]$ is determined

(implicitly) by the relation,

$$K(x) + \left[\frac{1}{\lambda}(\mathcal{X}(x) - \mu^2) - \frac{1}{2}J \circ G(\ ,x)G(x,\ ) \circ J + \frac{i}{2}G(x,x)\right]_{\mathcal{X}=\mathcal{X}_s} = 0, \qquad (2.8)$$

and the second derivative of the exponent in (2.6) with respect to $\mathcal{X}(y)$ is $-iND^{-1}$ with

$$\begin{aligned}D^{-1}[J,K](x,y) &\equiv -\frac{1}{\lambda}\delta^4(x,y)\\ &\quad - \left[J \circ G(\ ,x)G(x,y)G(y,\ ) \circ J - \frac{i}{2}G(x,y)G(y,x)\right]_{\mathcal{X}=\mathcal{X}_s}.\end{aligned}$$
(2.9)

Hence the result of the Gaussian stationary phase integration over $\mathcal{X}$ is

$$\begin{aligned}W[J,K] &\equiv W^{(0)} + \frac{1}{N}W^{(1)} + \frac{1}{N^2}W^{(2)} + \ldots\\ &= \frac{1}{\lambda}\mathcal{X}_s \circ (\frac{\mathcal{X}_s}{2} - \mu^2) + K \circ \mathcal{X}_s + \frac{1}{2}J \circ G[\mathcal{X}_s] \circ J + \frac{i}{2}\,\mathrm{Tr}\ln G^{-1}[\mathcal{X}_s]\\ &\quad + \frac{1}{2N}\,\mathrm{Tr}\ln D^{-1}[J,K] + \mathcal{O}\left(\frac{1}{N^2}\right)\end{aligned}$$
(2.10)

where $\mathcal{X}_s$ is to be viewed as a function of the sources $J$ and $K$ through eq. (2.8) above, and order $1/N^2$ terms have been dropped. By expanding the field $\mathcal{X}$ in a Taylor series about $\mathcal{X}_s$ in the original integral in (2.6) it is straightforward to derive the corrections to $W[J,K]$ appearing in (2.10) to any desired order in $1/N$. Hence the approximation scheme is systematic and controlled, and respects all the invariances of the original Lagrangian $L$ or $L'$ [16].

The mean field expectation values in the presence of the external sources are now given by the variations,

$$\overline{\phi}(x) \equiv \frac{\delta W}{\delta J(x)} \quad ; \quad \overline{\mathcal{X}}(x) \equiv \frac{\delta W}{\delta K(x)}. \qquad (2.11)$$

Note that the mean field $\overline{\mathcal{X}}$ differs from the stationary phase point of the Gaussian integral at order $1/N$, namely

$$\overline{\mathcal{X}} = \mathcal{X}_s + \frac{1}{N}\frac{\delta W^{(1)}}{\delta \mathcal{X}_s} \circ \frac{\delta \mathcal{X}_s}{\delta K} + \mathcal{O}\left(\frac{1}{N^2}\right). \qquad (2.12)$$

When no confusion with the integration variables in (2.5) is possible we shall omit the overline on the mean fields in the following to simplify the notation.

The effective action functional may be defined in terms of the mean fields (2.11) by a Legendre transformation in the usual way,

$$\mathcal{S}[\phi,\mathcal{X}] \equiv W - J \circ \phi - K \circ \mathcal{X}, \qquad (2.13)$$

where $J$ and $K$ are to be regarded here as functionals of the mean fields by inverting eqs. (2.11). Performing the inversion to order $1/N$, we find

$$J(x) = G^{-1}[\chi]\phi(x) + \frac{i}{N}\text{Tr}\{G(x,\ )\circ G \circ \phi \circ D\} -$$
$$\frac{1}{N}\frac{\delta W^{(1)}}{\delta \chi_s}\circ \frac{\delta \chi_s}{\delta J(x)}\Big|_{J=G^{-1}\circ\phi} + \mathcal{O}\left(\frac{1}{N^2}\right). \quad (2.14)$$

When this is substituted into (2.13) and account is taken of eqs. (2.8) and (2.12), most of the $1/N$ terms cancel and we arrive at the relatively simple result,

$$\mathcal{S}[\phi,\chi] = -\frac{1}{2}\phi\circ G^{-1}[\chi]\circ\phi + \frac{1}{\lambda}\chi\circ(\frac{\chi}{2}-\mu^2) + \frac{i}{2}\text{Tr}\ln G^{-1}[\chi] + \frac{i}{2N}\text{Tr}\ln D^{-1}[\phi,\chi], \quad (2.15)$$

with

$$D^{-1}[\phi,\chi](x,y) = -\frac{1}{\lambda}\delta^4(x,y) - \phi(x)G[\chi](x,y)\phi(y) + \frac{i}{2}G[\chi](x,y)G[\chi](y,x), \quad (2.16)$$

correct to order $1/N$. By differentiating the definition of the mean fields in (2.11) and using the form of the generating functional (2.10) and the stationarity condition (2.8) again, it is easy to check that $G[\chi]$ and $(1/N)D[\phi,\chi]$ are precisely the lowest order connected two-point functions of the fluctuating quantum fields, propagating in background mean fields specified by $(\phi,\chi)$. In the following the explicit functional dependence of $G[\chi]$ and $D[\phi,\chi]$ on the mean fields will be suppressed, and we adopt the simpler notation $G(x,y)$ and $D(x,y)$ hereafter.

From the explicit factor of $1/N$ multiplying the two-point function of the $\chi$ field and the last term in the effective action (2.15) it is clear that the fluctuations of the composite $\chi$ field enter the discussion at one higher order of $1/N$ than the original scalar $\phi$ field, whose fluctuations couple to the mean fields already at lowest order through the $\text{Tr}\ln G^{-1}$ term in the effective action. The fact that $\chi$ is an auxilliary field introduced into the discussion only for convenience and not a true propagating degree of freedom is reflected in the fact that there is no differential kinetic operator appearing in the expression for $D^{-1}$ (unlike in the definition of $G^{-1}$). The $\chi$ field has no independent dynamics of its own, and its fluctuations are determined by those of the $\phi$ field. The last term in (2.16), namely

$$\Pi[\chi] \equiv -\frac{i}{2}G[\chi]G[\chi] \quad (2.17)$$

is the one loop vacuum polarization of the $\chi$ field due to its interaction with the fluctuating $\phi$ field. It is included at the same order as the point vertex $1/\lambda$ in the large $N$ method because all loops of the $\phi$ field are exhanced by a factor of $N$ (for the $N$ identical $\phi$ fields) relative to what one would expect in an ordinary perturbative loop expansion. Finally we remark that

$$\Sigma(x,y) \equiv iG(x,y)D(y,x) \quad (2.18)$$

carries the interpretation of the one loop $\phi$ self-energy, as may be seen by calculating the $\phi$ inverse propagator function to order $1/N$,

$$\mathcal{G}^{-1}[\chi] = G^{-1} + \frac{1}{N}\Sigma + \mathcal{O}\left(\frac{1}{N^2}\right). \qquad (2.19)$$

The functional form of the quantum effective action (2.15) is the starting point for the analysis of the dynamics of the theory in the large $N$ expansion. By differentiating $\mathcal{S}$ with respect to the mean fields we obtain their equations of motion. In the absence of external sources these read:

$$(-\Box + \chi(x))\phi(x) + \frac{1}{N}\int d^4y\, \Sigma(x,y)\phi(y) = 0, \qquad (2.20)$$

and

$$\chi(x) = \mu^2 + \frac{\lambda}{2}\phi^2(x) - \frac{i\lambda}{2}G(x,x) + \frac{i\lambda}{2N}\int d^4x_1 \int d^4x_2\, G(x,x_1)\tilde{\Sigma}(x_1,x_2)G(x,x_2) \qquad (2.21)$$

with

$$\begin{aligned}\tilde{\Sigma}(x_1,x_2) &\equiv \Sigma(x_1,x_2) - \phi(x_1)D(x_1,x_2)\phi(x_2) \\ &= D(x_1,x_2)[iG(x_1,x_2) - \phi(x_1)\phi(x_2)]\end{aligned} \qquad (2.22)$$

The equation (2.21) for $\chi$ will be recognized as just the expectation value of the (operator) definition of $\chi$, eq. (2.4) computed to order $1/N$.

Since the original Lagrangian (2.1) possesses the symmetry $\phi \to -\phi$, eq. (2.20) is homogeneous in the $\phi$ mean field and always admits the solution $\phi = 0$. In this form it is suitable for study of second order phase transitions. If terms breaking the $\phi \to -\phi$ discrete symmetry are added (for example by retaining the linear source term $J \circ \phi$), then first order transitions may be studied as well. Both equations for the $\phi$ mean field and the $\chi$ mean field contain non-local self-energy effects, and are therefore *integro*-differential equations. In the case of the auxilliary $\chi$ field the differential term is absent, and eq. (2.21) is an equation of constraint (or gap equation) rather than a true propagating equation of motion for an independent degree of freedom.

In this derivation of the equations of motion for the mean fields from the quantum effective action (2.15) we have encountered the two-point functions $G$ and $D$. The equations of motion should be solved concurrently with those for the two-point functions obtained by inverting eqs. (2.2) and (2.16) respectively. Since there is no unique inverse of these relations, the question of *which* propagator function(s) should be chosen presents itself. If the standard Feynman propagator functions are substituted uncritically into eqs. (2.20) and (2.21), we find that the equations are both *complex* and *acausal*, in the sense that the integrations in (2.20) and (2.21) have support in regions of spacetime that are spacelike with respect to $x$. The absence of a well-posed causal initial value problem and the complex valuedness of the field equations both signal that this choice of Feynman boundary conditions for the Green's functions is not the correct one. The reason is that we do not wish to consider off-diagonal $\langle out|in\rangle$ matix elements of field operators but diagonal $\langle in|in\rangle$ expectation values. If the field operators are Hermitian, such diagonal matrix elements must be real. If one's next thought is to try purely retarded propagators in the equations of motion (2.20) and (2.21), the result will be no better. The equations are intrinsically non-linear, so that no one simple choice of particular solutions to the linear equations $G^{-1} \circ G = 1$ and $D^{-1} \circ D = 1$ for the propagator functions will yield real, causal equations of motion for the mean fields. This situation is familiar in non-relativistic condensed matter applications, and is the reason that a more complete method for deriving the correct boundary conditions as well as the equations for the mean fields must be introduced. Having pointed out the generic non-local structure of the field equations following from the quantum effective action, we defer the discussion of the Schwinger-Keldysh CTP formalism which is precisely one such method until the next section, preferring first to carry out the derivation of the large $N$ equations of motion in another interesting field theory, quantum electrodynamics.

For QED with $N$ identical charged fermion fields (flavors) we begin with the Lagrangian

$$L = -\sum_{i=1}^{N} \int d^4x \; \overline{\Psi}_i G^{-1}[A]\Psi_i - \frac{N}{4e^2} \int d^4x \; F_{\mu\nu}F^{\mu\nu} \qquad (2.23)$$

where anti-symmetrization with respect to the Dirac field operators $\Psi$ and $\overline{\Psi}$ is understood and

$$G^{-1}[A] = i\frac{\gamma^\mu}{2}(\vec{\partial}_\mu - \overleftarrow{\partial}_\mu) + \gamma^\mu A_\mu + im \; . \qquad (2.24)$$

The Dirac matrices here obey

$$\gamma^\mu\gamma^\nu + \gamma^\nu\gamma^\mu = 2g^{\mu\nu} = 2 \; \text{diag}(-,+,+,+) \qquad (2.25)$$

so that $g^0 = -\gamma_0$ is anti-hermitian and

$$\overline{\Psi} \equiv \Psi^\dagger \gamma^0 \; . \qquad (2.26)$$

Introducing external sources for the gauge potential and Dirac fields, we define the generating functional

$$Z[J,\eta,\overline{\eta}] \equiv \exp(iNW[J,K]) \;\equiv\; \int [\mathcal{D}A_\mu]' \prod_{i=1}^{N} \int [d\Psi_i][d\overline{\Psi}_i] \exp\left\{i \int d^4x \; L[A,\Psi,\overline{\Psi}]\right\}$$
$$\times \exp\left\{iNJ \circ A - \overline{\eta} \circ \Psi - \overline{\Psi} \circ \eta\right\}, \qquad (2.27)$$

where the prime on the gauge field integration measure denotes that we should integrate only over distinct gauge invariant configurations (or equivalently, fix the gauge).

Performing the Gaussian integration over the anti-commuting Dirac fields, and rescaling the Grassman valued sources $\eta \to \sqrt{N}\eta$ so that we can drop the sums over $i = 1, \ldots, N$ as in the previous scalar case, we obtain

$$\begin{aligned} Z[J, \eta, \overline{\eta}] &= \int [\mathcal{D}A_\mu]' \exp\left\{iN \int d^4x \, \frac{1}{2e^2} A_\mu \left(g^{\mu\nu}\Box - \partial^\mu \partial^\nu\right) A_\nu\right\} \\ &\quad \times \exp\left\{N\,\mathrm{Tr}\ln G^{-1} - N\overline{\eta} \circ G[A] \circ \eta + iNJ \circ A\right\}. \end{aligned} \quad (2.28)$$

As in the previous discussion we have defined the sources and coupling with the correct powers of $N$ to justify performing the remaining functional integration over the electromagnetic potential by the stationary phase method. The stationary phase value $A^s_\mu[J, \eta, \overline{\eta}]$ is fixed by

$$\frac{1}{e^2}\left(g^{\mu\nu}\Box - \partial^\mu \partial^\nu\right) A^s_\nu(x) = i\,\mathrm{tr}\left\{G[A^s](x,x)\gamma^\mu\right\} + i\overline{\eta} \circ G[A^s](\,,x)\gamma^\mu G[A^s](x,\,) \circ \eta - J^\mu(x) \quad (2.29)$$

where tr denotes the Dirac matrix trace only (without integration over spacetime coordinates). The second derivative of the exponent in (2.28) at its stationary point is $-iN\,\mathrm{Tr}\,D^{-1}$ where

$$\begin{aligned} D^{-1}[J, \eta, \overline{\eta}](x,y)^{\mu\nu} &= -\frac{1}{e^2}\left(g^{\mu\nu}\Box - \partial^\mu \partial^\nu\right)\delta^4(x,y) \\ &\quad - i\,\mathrm{tr}\left\{G(y,x)\gamma^\mu G(x,y)\gamma^\nu + 2\overline{\eta} \circ G(\,x)\gamma^\mu G(x,y)\gamma^\nu G(y,\,) \circ \eta\right\} \end{aligned} \quad (2.30)$$

and $A = A_s[J, \eta, \overline{\eta}]$ from the eq. (2.29) and symmetrization with respect to interchange of the pair of spacetime labels $(x, \mu)$ with $(y, \nu)$ is understood. Thus, the result of the stationary phase evaluation of (2.27) and (2.28) in the QED case is

$$\begin{aligned} W[J, K] &= -A^s \circ d^{-1} \circ A^s + J \circ A^s + i\overline{\eta} \circ G[A^s] \circ \eta \\ &\quad - i\,\mathrm{Tr}\ln G^{-1}[A^s] + \frac{i}{2N}\mathrm{Tr}\ln D^{-1}[A^s], \end{aligned} \quad (2.31)$$

where order $1/N^2$ terms have been dropped, and

$$d^{-1}(x,y)^{\mu\nu} \equiv -\frac{1}{e^2}\left(g^{\mu\nu}\Box - \partial^\mu \partial^\nu\right)\delta^4(x,y) \quad (2.32)$$

is the differential operator from the classical action.

The mean fields and quantum effective action are defined now by the analogs of eqs. (2.11), and (2.13), after solving for the sources $J, \eta, \overline{\eta}$ in terms of the mean fields. Omitting the details which are quite analogous to the scalar case, the result of this Legendre transformation is simply

$$\mathcal{S}[A] = -\frac{1}{2} A \circ d^{-1} \circ A - i\,\mathrm{Tr}\ln G^{-1}[A] + \frac{i}{2N}\mathrm{Tr}\ln D^{-1}[A], \quad (2.33)$$

in the case of zero mean value for the Dirac field. The photon inverse propagator in the last term is given by

$$D^{-1}[A](x,y)^{\mu\nu} = (d^{-1} + \Pi[A])(x,y)^{\mu\nu} \qquad (2.34)$$

with

$$\Pi[A](x,y)^{\mu\nu} \equiv -i\text{tr}\left\{\gamma^{\mu}G[A](x,y)\gamma^{\nu}G[A](y,x)\right\} \qquad (2.35)$$

the polarization tensor in the presence of the mean potential $A$. The inverse propagator cannot be inverted without fixing a gauge, which may be done by a variety of standard methods. In the case of a non-abelian gauge symmetry, the gauge fixing introduces ghosts which will also contribute to the quantum effective action at order $1/N$, and whose contribution is essential to obtaining gauge invariant results. In the abelian QED case the ghosts are independent of the mean potential (*i.e.*, they decouple) and therefore may be omitted from the effective action (2.33). Of course, the non-abelian case is of great interest for describing the non-equilibrium evolution of the quark gluon plasma from first principles of QCD, and will be discussed in detail in a future publication.

The integro-differential equation for the mean potential is obtained by varying the effective action. In this variation will appear

$$\begin{aligned}\frac{\delta \Pi[A](x_1,x_2)^{\alpha\beta}}{\delta A_\mu(x)} &= i\text{tr}\left[\gamma^\alpha G(x_1,x)\gamma^\mu G(x,x_2)\gamma^\beta G(x_2,x_1)\right] \\ &+ i\text{tr}\left[\gamma^\alpha G(x_1,x_2)\gamma^\beta G(x_2,x)\gamma^\mu G(x,x_1)\right] ,\end{aligned} \qquad (2.36)$$

so that the equations of motion for the mean potential read

$$\begin{aligned}\partial_\nu F^{\mu\nu}(x) &= -ie^2\text{tr}\left\{\gamma^\mu G[A](x,x)\right\} + \frac{ie^2}{2N}\text{Tr}\left\{D[A] \circ \frac{\delta \Pi[A]}{\delta A_\mu(x)}\right\} \\ &= -ie^2\text{tr}\left\{\gamma^\mu G(x,x)\right\} \\ &\quad + \frac{ie^2}{N}\int d^4x_1 \int d^4x_2\, \text{tr}\left\{\gamma^\mu G(x,x_1)\Sigma(x_1,x_2)G(x_2,x)\right\} \end{aligned} \qquad (2.37)$$

with

$$\Sigma(x_1,x_2) \equiv i\gamma^\mu G(x_1,x_2)\gamma^\nu D_{\nu\mu}(x_2,x_1) \qquad (2.38)$$

the fermion self-energy. The current expectation value to order $1/N$ is represented pictorially by the graphs in Fig. 1.

To leading order in $1/N$ the mean field equations are just the semi-classical Maxwell equations, obtained by replacing the electric current operator of the Dirac field by its expectation value. This leading order semi-classical equation already contains the dynamical reaction of $e^+e^-$ pairs created by a non-zero electric field (the

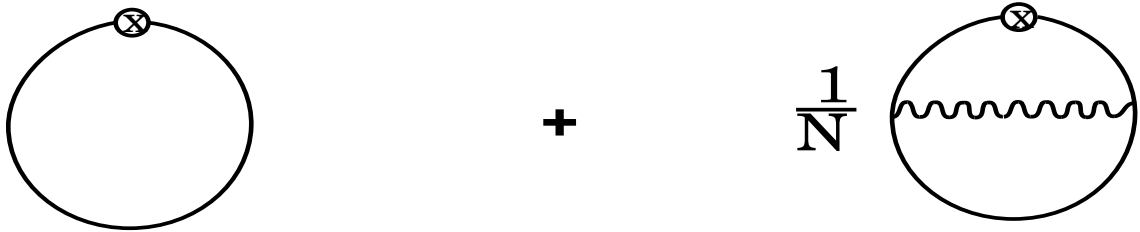

Figure 1: First two graphs in the $1/N$ expansion contributing to the induced current that governs the back reaction on the electric field.

Schwinger mechanism) back on the electric field itself, and has been studied in previous publications [10]. However, at leading order in $1/N$ the created pairs can interact only through the mean field, not directly with each other. The order $1/N$ term with the fermion self-energy $\Sigma$ contains the quantum Compton scattering, bremmstrahlung and Coulomb interaction effects of these particles on each other and their backreaction on the self-consistent mean field. Clearly these processes are essential mechanisms for the approach to equilibrium and must be included in any realistic transport theory of a relativistic $e^+e^-$ plasma. Higher order processes (such as multiple scattering) can be included by working to higher orders in the $1/N$ expansion in a straightforward way.

Having derived the integro-differential equations for the mean fields in both scalar $\lambda\Phi^4$ theory and electrodynamics to order $1/N$, we turn now to the Schwinger-Keldysh closed time path formulation of the effective action, in order to determine the correct propagator functions needed to obtain a causal (and real) solution to these equations.

## 3  The Schwinger-Keldysh Closed Time Path Formalism

The conventional path integral formalism used freely in the preceeding section defines transition elements between states at one time, $t$ (usually taken to be in the infinite past) to states at another time $t'$ (in the distant future). If the class of paths is restricted to be the vacuum configuration at both of its endpoints, then the two states are the $|in\rangle$ and $\langle out|$ vacuum states of scattering theory respectively. The functional $Z[J,K]$ of eq. (2.5) is the transition matrix element

$$Z[J,K](t,t') = \langle out, t'|in, t\rangle_{J,K} \tag{3.1}$$

in the presence of the external sources $J$ and $K$.

By varying with respect to the external sources we obtain matrix elements of the Heisenberg field operators between the $|in\rangle$ and $\langle out|$ states. For this reason we may refer to the conventional formulation of the generating functional $Z$ as the "in-out" formalism. The time-ordered Green's functions obtained in this way necessarily obey Feynman boundary conditions, and these are the appropriate ones for the calculation of transition probabilities and cross sections between the $|in\rangle$ and $\langle out|$ states. On the other hand the off-diagonal transition matrix elements of the in-out formalism are completely inappropriate if what we wish to consider is the time evolution of physical observables from a given set of initial conditions. As we have remarked the in-out matrix elements are neither real, nor are their equations of motion causal at first order in $1/N$, where direct self interactions between the fields appear for the first time. What we require is a generating functional for *diagonal* matrix elements of field operators with a corresponding modification of the Feynman boundary conditions on Green's functions to ensure causal time evolution. This "in-in" formalism was developed more than thirty years ago by Schwinger and later by Keldysh, and is called the closed time path (CTP) method [12].

The basic idea of the CTP formalism is to take a diagonal matrix element of the system at a given time $t = 0$ and insert a complete set of states into this matrix element at a different (later) time $t'$. In this way one can express the original fixed time matrix element as a product of transition matrix elements from 0 to $t'$ and the time reversed (complex conjugate) matrix element from $t'$ to 0. Since each term in this product is a transition matrix element of the usual or time reversed kind, standard path integral representations for each may be introduced. If the same external source operates in the forward evolution as the backward one, then the two matrix elements are precisely complex conjugates of each other, all dependence on the source drops out and nothing has been gained. However, if the forward time evolution takes place in the presence of one source $J_+$ but the reversed time evolution takes place in the presence of a *different* source $J_-$, then the resulting functional is precisely the generating functional we seek. Indeed (setting $K = 0$ and $N = 1$ here for simplicity),

$$\begin{aligned} Z_{in}[J_+, J_-] &\equiv \int [\mathcal{D}\Psi] \langle in|\psi\rangle_{J_-} \langle\psi|in\rangle_{J_+} \\ &= \int [\mathcal{D}\Psi] \langle in|\mathcal{T}^* exp\left[-i\int_0^{t'} dt d^3\vec{x} J_-(x)\phi(x)\right] |\Psi, t'\rangle \\ &\quad \times \langle\Psi, t'|\mathcal{T} \exp\left[i\int_0^{t'} dt d^3\vec{x} J_+(x)\phi(x)\right] |in\rangle \end{aligned} \qquad (3.2)$$

so that, for example,

$$\left.\frac{\delta W_{in}[J_+, J_-]}{\delta J_+(x)}\right|_{J_+ = J_- 0} = -\left.\frac{\delta W_{in}[J_+, J_-]}{\delta J_-(x)}\right|_{J_+ = J_- = 0} = \langle in|\phi(x)|in\rangle \qquad (3.3)$$

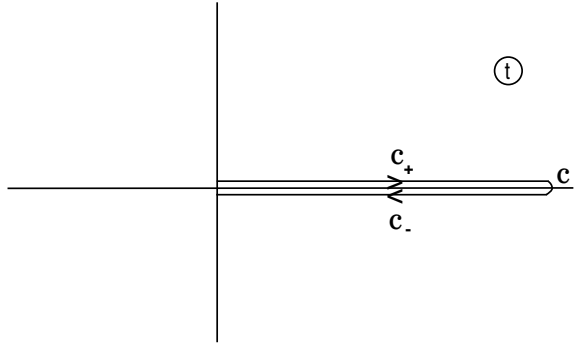

Figure 2: Complex time contour $\mathcal{C}$ for the closed time path propagators.

is a true field expectation value in the given time-independent Heisenberg state $|in\rangle$. Since the time ordering in eq. (3.2) is forward (denoted by $\mathcal{T}$) along the time path from 0 to $t'$ in the second transition matrix element, but backward (denoted by $\mathcal{T}^*$) along the path from $t'$ to 0 in the first matrix element, this generating functional receives the name of the closed time path generating functional. If we deform the backward and forward directed segments of the path slightly in opposite directions in the complex $t$ plane, the symbol $\mathcal{T}_\mathcal{C}$ may be introduced for path ordering along the full closed time contour, $\mathcal{C}$ depicted in Fig. 2.

This deformation of the path corresponds precisely to opposite $i\epsilon$ prescriptions along the forward and backward directed segments, which we shall denote by $\mathcal{C}_\pm$ respectively in the following.

The doubling of sources, fields and integration contours in the CTP formalism may seem artificial, but in fact it appears naturally as soon as one discusses the time evolution not of states in Hilbert space but of density matrices. Then it is clear that whereas $|\ \rangle$ ket states evolve with Hamiltonian $H$, the conjugate $\langle\ |$ bra states evolve with $-H$, and the evolution of the density matrix requires both. Hence a doubling of all sources and fields in the functional integral representation of its time evolution kernel is necessary. Indeed, it is easy to generalize the functional in (3.2) to the case of an arbitrary initial density matrix $\rho$, by defining

$$\begin{aligned}
Z[J_+, J_-, \rho] &\equiv \mathrm{Tr}\left\{\rho\left(\mathcal{T}^*\exp\left[-i\int_0^{t'} dt d^3\vec{x} J_-(x)\phi(x)\right]\right)\right.\\
&\qquad \left.\times\left(\mathcal{T}\exp\left[i\int_0^{t'} dt d^3\vec{x} J_+(x)\phi(x)\right]\right)\right\}\\
&= \int[\mathcal{D}\varphi][\mathcal{D}\varphi'][\mathcal{D}\psi]\,\langle\varphi|\rho|\varphi'\rangle\,\langle\varphi'|\mathcal{T}^*exp\left[-i\int_0^{t'} dt d^3\vec{x} J_-(x)\phi(x)\right]|\psi\rangle\\
&\qquad \times\langle\psi|\mathcal{T}exp\left[i\int_0^{t'} dt d^3\vec{x} J_+(x)\phi(x)\right]|\varphi\rangle\ . \hspace{2em} (3.4)
\end{aligned}$$

Variations of this generating function will yield Green's functions in the state specified by the initial density matrix, *i.e.*, expressions of the form,

$$\mathrm{Tr}\{\rho \phi(x_1)\phi(x_2)\phi(x_3)...\}. \tag{3.5}$$

Introducing the path integral representation for each transition matrix element in eq. (3.4) results in the expression,

$$\begin{aligned}
Z[J_+, J_-, \rho] &= \int [\mathcal{D}\varphi][\mathcal{D}\varphi'] \langle\varphi|\rho|\varphi'\rangle \int [\mathcal{D}\psi] \int_\varphi^\psi [\mathcal{D}\phi_+] \int_{\varphi'}^\psi [\mathcal{D}\phi_-] \times \\
&\quad \exp\left[i\int_0^\infty dt d^3\vec{x}\left(L[\phi_+] - L[\phi_-] + J_+\phi_+ - J_-\phi_-\right)\right],
\end{aligned} \tag{3.6}$$

where $L$ is the classical Lagrangian functional, and we have taken the arbitrary future time at which the time path closes $t' \to \infty$.

The double path integral over the fields $\phi_+$ and $\phi_-$ in (3.6) suggests that we introduce a two component contravariant vector of field variables by

$$\phi^a = \begin{pmatrix} \phi_+ \\ \phi_- \end{pmatrix}; \quad a = 1, 2 \tag{3.7}$$

with a corresponding two component source vector,

$$J^a = \begin{pmatrix} J_+ \\ J_- \end{pmatrix}; \quad a = 1, 2. \tag{3.8}$$

Because of the minus signs in the exponent of (3.6), it is necessary to raise and lower indices in this vector space with a $2 \times 2$ matrix with indefinite signature, namely

$$c_{ab} = diag(+1, -1) = c^{ab} \tag{3.9}$$

so that, for example

$$J^a c_{ab} \Phi^b = J_+\phi_+ - J_-\phi_-. \tag{3.10}$$

These definitions imply that the correlation functions of the theory will exhibit a matrix structure in the $2 \times 2$ space. For instance, the matrix of connected two point functions in the CTP space is

$$G^{ab}(x,y) = \left.\frac{\delta^2 W}{\delta J_a(x)\delta J_b(y)}\right|_{J=0}. \tag{3.11}$$

Explicitly, the components of this $2 \times 2$ matrix are

$$\begin{aligned}
G^{21}(x,y) &\equiv G_>(x,y) = i\mathrm{Tr}\{\rho\ \Phi(x)\overline{\Phi}(y)\}_{con}, \\
G^{12}(x,y) &\equiv G_<(x,y) = \pm i\mathrm{Tr}\{\rho\ \overline{\Phi}(y)\Phi(x)\}_{con}, \\
G^{11}(x,y) &= i\mathrm{Tr}\left\{\rho\ T[\Phi(x)\overline{\Phi}(y)]\right\}_{con} = \theta(x,y)G_>(x,y) + \theta(y,x)G_<(x,y), \\
G^{22}(x,y) &= i\mathrm{Tr}\left\{\rho\ T^*[\Phi(x)\overline{\Phi}(y)]\right\}_{con} = \theta(y,x)G_>(x,y) + \theta(x,y)G_<(x,y),
\end{aligned} \tag{3.12}$$

where the ± refers to Bose or Fermi fields respectively, and $\overline{\Phi} \equiv \Phi^\dagger$ for bosons but $\overline{\Psi} \equiv \Psi^\dagger \gamma^0$ for Dirac fermions. Notice that $G_>(x,y) = G_<(x,y)$ if $x$ and $y$ are spacelike, and therefore

$$G^{11}(x,y) = G^{22}(x,y) \qquad \text{for } (x,y) \text{ spacelike.} \tag{3.13}$$

and in particular on the spacelike surface $t_x = t_y$. This will be fully consistent with the expressions for time ordered and anti-time ordered propagator functions in (3.13), provided we extend the usual definition of the discontinuous $\theta$ function so that

$$\theta(t,t') + \theta(t',t) = 1 \qquad \text{for all } t,t' \tag{3.14}$$

so that in particular,

$$\theta(t,t) = \frac{1}{2}. \tag{3.15}$$

The $2 \times 2$ matrix notation has been discussed extensively in the literature [12]. However, the development of the CTP formalism is cleaner, both conceptually and notationally, by returning to the definition of the generating functional (3.4), and using the composition rule for transition amplitudes along the closed time contour in the complex plane. Then we may dispense with the $2 \times 2$ matrix notation altogether, and write simply

$$\int [\mathcal{D}\psi] \langle \varphi' | \mathcal{T}^* \exp\left[-i \int_0^\infty dt d^3\vec{x}\, J_-(x)\phi(x)\right] |\psi\rangle \langle \psi | \mathcal{T} \exp\left[i \int_0^\infty dt d^3\vec{x}\, J_+(x)\phi(x)\right] |\varphi\rangle$$

$$= \langle \varphi' | \mathcal{T}_{\mathcal{C}} \exp\left[i \int_{\mathcal{C}} dt d^3\vec{x}\, J(x)\phi(x)\right] |\varphi\rangle \tag{3.16}$$

so that (3.4) may be rewritten more concisely in the CTP complex path ordered form,

$$\begin{aligned} Z_{\mathcal{C}}[J,\rho] &= \text{Tr}\left\{\rho \left(\mathcal{T}_{\mathcal{C}} \exp\left[i \int_{\mathcal{C}} dt d^3\vec{x} J(x)\phi(x)\right]\right)\right\} \\ &= \int [\mathcal{D}\varphi^1] \int [\mathcal{D}\varphi^2]\, \langle \varphi^1 | \rho | \varphi^2 \rangle \int_{\varphi^1}^{\varphi^2} [\mathcal{D}\phi] \exp\left[i \int_{\mathcal{C}} dt d^3\vec{x}\, (L[\phi] + J\phi)\right]. \end{aligned} \tag{3.17}$$

The advantage of this form is that it is identical in structure to the usual expression for the generating functional in the more familiar in-out formalism, with the only difference of path ordering according to the complex time contour $\mathcal{C}$ replacing the ordinary time ordering prescription along only $\mathcal{C}_+$. Hence, all the functional formalism of the previous section may be taken over line for line, with only this modification of complex path ordering in the time integrations. For example, the propagator function becomes

$$\begin{aligned} G(x,y) &= \theta_{\mathcal{C}}(t_x,t_y) G_>(x,y) + \theta_{\mathcal{C}}(t_y,t_x) G_<(x,y) \\ &\equiv \theta_{\mathcal{C}}(t_x,t_y) G^{21}(x,y) + \theta_{\mathcal{C}}(t_y,t_x) G^{12}(x,y) \end{aligned} \tag{3.18}$$

where $\theta_c$ is the CTP complex contour ordered theta function defined by

$$\theta_c(t,t') \equiv \begin{cases} \theta(t,t') & \text{for } t,t' \text{ both on } \mathcal{C}_+ \\ \theta(t',t) & \text{for } t,t' \text{ both on } \mathcal{C}_- \\ 1 & \text{for } t \text{ on } \mathcal{C}_- \text{ , } t' \text{ on } \mathcal{C}_+ \\ 0 & \text{for } t \text{ on } \mathcal{C}_+ \text{ , } t' \text{ on } \mathcal{C}_- \end{cases} \quad (3.19)$$

With this definition of $G(x,y)$ on the closed time contour, the Feynman rules are the ordinary ones, and matrix indices are not required. In integrating over the second half of the contour $\mathcal{C}_-$ we have only to remember to multiply by an overall negative sign to take account of the opposite direction of integration, according to the rule,

$$\int_{\mathcal{C}} dt = \int_{0\ \mathcal{C}_+}^{\infty} dt - \int_{0\ \mathcal{C}_-}^{\infty} dt \ . \quad (3.20)$$

A second simplification is possible in the form of the generating functional of (3.17), if we recognize that it is always possible to express the matrix elements of the density matrix as an exponential of a polynomial in the fields [5]:

$$\langle \varphi^1 | \rho | \varphi^2 \rangle = \exp\left[ R + \int d^3\vec{x}\, R_a(\vec{x}) \varphi^a(\vec{x}) + \int d^3\vec{x} d^3\vec{y}\, R_{ab}(\vec{x},\vec{y}) \varphi^a(\vec{x}) \varphi^b(\vec{y}) + \ldots \right] \ . \quad (3.21)$$

Since any density matrix can be expressed in this form, there is no loss of generality involved in expressing $\rho$ as an exponential. If we add this exponent to that of the action in (3.17), and integrate over the two endpoints of the closed time path $\varphi^1$ and $\varphi^2$, then the only effect of the non-trivial density matrix $\rho$ is to introduce source terms into the path integral for $Z_{\mathcal{C}}[J,\rho]$ with support *only* at the endpoints. This means that the density matrix can only influence the boundary conditions on the path integral at $t = 0$, where the various coefficient functions $R_a$, $R_{ab}$, *etc.* have the simple interpretations of initial conditions on the one-point (mean field), two-point (propagator), functions *etc.* It is clear that the equations of motion for $t \neq 0$ are not influenced by the presence of these terms at $t = 0$. In the special case that the initial density matrix describes a thermal state, $\rho_\beta = \exp\{-\beta H\}$ then the trace over $\rho_\beta$ may be represented as an additional functional integration over fields along the purely imaginary contour from $t = -i\beta$ to $t = 0$ traversed before $\mathcal{C}_-$ in Fig. 2. In this way the Feynman rules for real time thermal Green's functions are obtained [13]. Since we consider general nonequilibrium initial conditions here we have only the general expression for the initial $\rho$ above and no contour along the negative imaginary axis in Fig. 2.

To summarize, we may take over all the results of the previous section on the generating functionals, effective actions, and equations of motion of scalar $\Phi^4$ theory and QED, provided only that we

1. substitute the CTP path ordered Green's function(s) (3.18) for the ordinary Feynman propagators in internal lines;

2. integrate over the full closed time contour, $\mathcal{C}$, according to (3.20); and

3. satisfy the conditions at $t = 0$ corresponding to the initial density matrix $\rho$.

# 4 Causal Evolution Equations in $\Phi^4$ Theory and QED

To show that these three modifications of ordinary in-out formalism do lead to a well-posed initial value problem for quantum field theory with real and causal equations of motion for the mean fields and their Green's functions, let us reconsider eqs. (2.20) and (2.21), Section 2. First, the self-energy has the CTP ordered form,

$$\Sigma(x,y) = i\theta_\mathcal{C}(t_x, t_y) G_>(x,y) D_>(x,y) + i\theta_\mathcal{C}(t_y, t_x) G_<(x,y) D_<(x,y) \quad (4.1)$$

since both $G(x,y)$ and $D(x,y)$ have this form separately. Using the rule (3.20) for the time contour integration, we have

$$\begin{aligned}
\int_\mathcal{C} dt_y \Sigma(x,y)\phi(y) &= i\int_0^{t_x} dt_y \, G_>(x,y) D_>(x,y)\phi(y) + i\int_{t_x}^\infty dt_y \, G_<(x,y) D_<(x,y)\phi(y) \\
&\quad - i\int_0^\infty dt_y \, G_<(x,y) D_<(x,y)\phi(y) \\
&= i\int_0^{t_x} dt_y \, [G_>(x,y) D_>(x,y) - G_<(x,y) D_<(x,y)]\phi(y) \\
&= -2\int_0^{t_x} dt_y \, \mathcal{I}m \, [G_>(x,y) D_>(x,y)]\phi(y) \quad (4.2)
\end{aligned}$$

The minus sign in the second line is because $t_y$ is on the second branch of the contour, and the fact that

$$\begin{aligned}
[G_>(x,y)]^* &= -G_<(x,y) \\
[D_>(x,y)]^* &= -D_<(x,y)
\end{aligned} \quad (4.3)$$

for *real boson* fields has been used. Therefore, the equation of motion for the mean field, (2.20), is

$$(-\Box + \mathcal{X}(x))\phi(x) - \frac{2}{N}\int_0^{t_x} dt_y d^3\vec{y} \, \mathcal{I}m \, [G_>(x,y) D_>(x,y)]\phi(y) = 0 \,. \quad (4.4)$$

This equation is now explicitly both real and causal. Furthermore, it is not difficult to see that the cancellation of the acausal parts of the integration (when $t_y > t_x$ here) is completely general in the CTP method, since every internal vertex requires a time integration over the full path $\mathcal{C}$, and equal and opposite contributions will always come from the portions of the integration on $\mathcal{C}_+$ and $\mathcal{C}_-$ with $t_y > t_x$. One simply decomposes the time integration for each internal vertex $t_i$ into three segments, viz.,

$$\begin{aligned}(i) & \quad 0 < t_i < t, \text{ on } \mathcal{C}_+ \\ (ii) & \quad t < t_i < \infty, \text{ on } \mathcal{C}_+ \\ (iii) & \quad 0 < t_i < \infty, \text{ on } \mathcal{C}_-\end{aligned} \quad (4.5)$$

uses the definitions (3.18), (3.19), (3.20), and collects the non-cancelling terms. In this way we find that the $1/N$ term appearing in the eq. (2.21) for the $\chi$ mean field reduces to

$$\begin{aligned}i \int_\mathcal{C} dt_1 \int_\mathcal{C} dt_2 G(t,t_2) G(t_1,t) \widetilde{\Sigma}(t_1,t_2) &= i \int_0^t dt_1 \int_0^t dt_2 \big\{ G_>(t,t_2) G_<(t_1,t) \widetilde{\Sigma}(t_1,t_2) \\ &\quad - G_<(t,t_2) G_<(t_1,t) \widetilde{\Sigma}_<(t_1,t_2) \\ &\quad - G_>(t,t_2) G_>(t_1,t) \widetilde{\Sigma}_>(t_1,t_2) \\ &\quad - G_<(t,t_2) G_>(t_1,t) \widetilde{\Sigma}^*(t_1,t_2) \big\},\end{aligned} \quad (4.6)$$

where the last term is anti-time ordered since $t_1$ and $t_2$ are both on $\mathcal{C}_-$, and all spatial dependences have been suppressed. By now expanding in the possible orderings of $t_1$ and $t_2$, interchanging integration variables, and using

$$\begin{aligned}[G_{>,<}(x,y)]^* &= -G_{>,<}(y,x) \\ [D_{>,<}(x,y)]^* &= -D_{>,<}(y,x)\end{aligned} \quad (4.7)$$

which is true for real or complex Bose fields, we can write (4.6) in the form,

$$2\,\mathcal{I}m \int_0^t dt_1 \int_0^{t_1} dt_2 \big[G_>(t_1,t) - G_<(t_1,t)\big]\big[G_>(t,t_2) \widetilde{\Sigma}_>(t_1,t_2) - G_<(t,t_2) \widetilde{\Sigma}_<(t_1,t_2)\big] \quad (4.8)$$

which is manifestly real and causal. It is easy to see that this result is identical to that obtained in the matrix notation from

$$i \int_0^t dt_1 \int_0^t dt_2\, G^{1b}(t,t_2) G^{a1}(t_1,t) \widetilde{\Sigma}_{ab}(t_1,t_2) \quad (4.9)$$

provided indices are raised and lowered with the indefinite metric $c_{ab}$ defined in eq. (3.9). Since the causality and reality of the equations are assured on general grounds, the result of any calculation is obtained most rapidly by proceeding to this matrix form directly.

The effective Feynman rules for the CTP method in the large $N$ expansion may be summarized then as follows. Derive the effective action and equations of motion for the mean fields and propagator functions to a given order in $1/N$ in the usual in-out formalism, as in the previous section. In the corresponding graphical representation of these equations, replace the propagators with the $2 \times 2$ matrix propagator of the CTP method, raising and lowering all contracted indices at internal vertices with the metric $c_{ab}$, and fixing the "external" vertex of the mean field(s) to be of type 1. After all time orderings have been taken into account only the Wightman functions such as $G_> = G^{21}$ or $G_< = G^{12}$ will appear in the equations, which can be solved for concurrently with the corresponding mean fields in a well-posed real and causal intitial value problem.

For the $\Phi^4$ theory we have gone through the CTP procedure in some detail using the above manipulations. The final result for the equation of the $\phi$ mean field has been given already by (4.4). The corresponding causal $\mathcal{X}$ equation (2.21) is

$$\begin{aligned} \mathcal{X}(x) &= \mu^2 + \frac{\lambda}{2}\phi^2(x) - \frac{i\lambda}{4}[G_>(x,x) + G_<(x,x)] \\ &\quad + \frac{\lambda}{N}\, \mathcal{I}m \int_0^t dt_1 d^3\vec{x}_1 \int_0^{t_1} dt_2 d^3\vec{x}_2 \, [G_>(x,x_1) - G_<(x,x_1)] \\ &\quad \times \, \left[\tilde{\Sigma}_<(x_1,x_2)G_>(x_2,x) - \tilde{\Sigma}_>(x_1,x_2) - G_<(x_2,x)\right] \, , \end{aligned} \qquad (4.10)$$

with $\tilde{\Sigma}$ given by eq. (2.22) of the previous section.

Concurrently with the mean field equations we must solve for the two-point functions $G$ and $D$. If both have the causal form of (3.18) in the path ordered notation, then the Wightman functions $G_{>,<}$ satisfy the homogeneous equations,

$$\left(-\Box + \mathcal{X}(x)\right) G_{>,<}(x,x') = 0 \, , \qquad (4.11)$$

together with the initial conditions following from the initial density matrix $\rho$ as specified by eqs. (3.13). Because of the canonical equal time commutation relations the initial conditions on the first derivatives of $G_{>,<}$ must satisfy the constraint,

$$\frac{d}{dt_x}[G_>(x,y) - G_<(x,y)]_{t_x=t_y} = \delta^3(\vec{x}-\vec{y}) \qquad (4.12)$$

which guarantees that the inhomogeneous equation $G^{-1} \circ G = 1$ is satisfied by (3.18). There is no corresponding derivative condition for the $D$ propagator function since the operator $D^{-1}$ of (2.16) does not contain any time derivatives. Instead, the function $D$ satisfies the inhomogeneous integral equation,

$$\delta^4(x,y) = -\frac{1}{\lambda}D(x,y) - \int_{\mathcal{C}} d^4x_1 \tilde{\Pi}(x,x_1) D(x_1,y) \qquad (4.13)$$

with

$$\begin{aligned} \tilde{\Pi}(x,y) &\equiv \Pi(x,y) + \phi(x)G(x,y)\phi(y) \\ &= G(x,y)\left[-\frac{i}{2}G(y,x) + \phi(x)\phi(y)\right] \, . \end{aligned} \qquad (4.14)$$

The only way to satisfy the inhomogeneous equation (4.13) is for the $D_{>,<}$ functions themselves to contain a $\delta$ function term. This is simply a consequence of the fact that the $\chi$ field defined by eq. (2.4) is an auxilliary field, purely constrained in terms of $\phi$ by its definition with no conjugate momentum or independent dynamics of its own. Hence we seek a solution of (4.13) of the form,

$$D_{>,<}(x,y) = -\lambda \delta^4(x,y) + \widetilde{D}_{>,<}(x,y) \tag{4.15}$$

with $\widetilde{D}_{>,<}$ smooth functions satisfying

$$\begin{aligned}\widetilde{D}_{<,>}(x,y) &= \lambda^2 \widetilde{\Pi}_{<,>}(x,y) - \lambda \int_0^{t_x} d^4 x_1 [\widetilde{\Pi}_{>}(x,x_1) - \widetilde{\Pi}_{>}(x,x_1)] D_{<,>}(x_1,x) \\ &+ \lambda \int_0^{t_y} d^4 x_1\, \widetilde{\Pi}_{<,>}(x,x_1)[\widetilde{D}_{>}(x_1,y) - \widetilde{D}_{<}(x_1,y)]\,.\end{aligned} \tag{4.16}$$

Evidently the functions $\widetilde{D}_{>,<}$ are determined completely by the initial conditions and causal evolution of the $\phi$ field and its propagator functions $G_{>,<}$, consistent with $\chi$ being a fully constrained field with no independent dynamics.

This completes the derivation of the causal equations of motion for the initial value problem of nonequilibrium $\Phi^4$ theory to order $1/N$. The derivation of the corresponding equations for nonequilibrium electrodynamics proceeds in an exactly analogous manner with the only difference that the relation

$$[G_{>,<}(x,y)]^* = -\gamma^0 G_{>,<}(y,x)\gamma^0 \tag{4.17}$$

for complex Dirac fields replaces the first member of (4.7). The result is that the causal Maxwell equations of motion take the form,

$$\begin{aligned}\partial_\nu F^{\mu\nu}(x) &= \langle j^\mu(x)\rangle = -\frac{ie^2}{2}\mathrm{tr}\left\{\gamma^\mu[G_{>}(x,x) + G_{<}(x,x)]\right\} \\ &+ \frac{2e^2}{N}\int_0^t dt_1 d^3\vec{x}_1 \int_0^{t_1} dt_2 d^3\vec{x}_2 \mathcal{I}m\,\mathrm{tr}\Big\{\gamma^\mu\left[G_{>}(x,x_1) - G_{<}(x,x_1)\right] \\ &\times \left[\Sigma_{<}(x_1,x_2)G_{>}(x_2,x) - \Sigma_{>}(x_1,x_2)G_{<}(x_2,x)\right]\Big\}\,.\end{aligned} \tag{4.18}$$

The Wightman functions for the Dirac field satisfy

$$(\gamma^\mu \partial_\mu - i\gamma^\mu A_\mu + m) \equiv (\gamma^\mu \nabla_\mu + m)G_{>,<}(x,y) = 0, \tag{4.19}$$

together with the initial conditions implied by the first two members of eqs. (3.13), which satisfy the canonical equal time anticommutator condition,

$$-i\left[G_{>}(x,y) - G_{<}(x,y)\right]_{t_x=t_y} = \mathrm{Tr}\left\{\rho[\Psi(x),\overline{\Psi}(y)]_+\right\}_{t_x=t_y} = \gamma^0 \delta^3(\vec{x}-\vec{y})\,, \tag{4.20}$$

appropriate for Fermi-Dirac statistics.

For a complete initial value problem to order $1/N$, one needs also the two-point function of the Maxwell field obtained by inverting (2.34) subject to some gauge condition. The simplest way to impose the gauge condition is to treat it in a way similar to the constraint of the $\lambda\Phi^4$ theory, *i.e.*, we write

$$D_{\mu\nu}(x,y) = e^2 d_{\mu\nu}(x,y) + \widetilde{D}_{\mu\nu}(x,y) \qquad (4.21)$$

with $d_{\mu\nu}$ the inverse of the differential operator (2.32) of the *free* action in a definite gauge. The gauge fixing can be performed at the level of the free photon propagator once and for all, independently of the dynamical time evolution problem, and the non-trivial time evolution is contained entirely in $\widetilde{D}_{\mu\nu}$, which obeys

$$\left( g^{\mu\lambda}\Box - \partial^{\mu}\partial^{\lambda} \right) \widetilde{D}_{\lambda\nu}(x,y) = e^2 \int_{\mathcal{C}} d^4 x_1 \Pi^{\mu\lambda}(x,x_1) \left( e^2 d_{\lambda\nu}(x_1,y) + \widetilde{D}_{\lambda\nu}(x_1,y) \right) \; . \qquad (4.22)$$

Writing the propagator in terms of its time ordered structure along the contour $\mathcal{C}$ as in eq. (3.18) we obtain

$$\begin{aligned}
&\left( g^{\mu\lambda}\Box - \partial^{\mu}\partial^{\lambda} \right) \widetilde{D}_{\lambda\nu}^{>,<}(x,y) = \\
&-e^2 \int_0^{t_y} d^4 x_1 \, \Pi_{>,<}^{\mu\lambda}(x,x_1) \left[ e^2 d_{\lambda\nu}^{>}(x_1,y) + \widetilde{D}_{\lambda\nu}^{>}(x_1,y) - e^2 d_{\lambda\nu}^{<}(x_1,y) - \widetilde{D}_{\lambda\nu}^{<}(x_1,y) \right] \\
&+e^2 \int_0^{t_x} d^4 x_1 \left[ \Pi_{>}^{\mu\lambda}(x,x_1) - \Pi_{<}^{\mu\lambda}(x,x_1) \right] \left[ e^2 d_{\lambda\nu}^{>,<}(x_1,y) + \widetilde{D}_{\lambda\nu}^{>,<}(x_1,y) \right] \; , \qquad (4.23)
\end{aligned}$$

which is explicitly causal. We remark that a particularly useful gauge choice for practical implementation of the initial value problem on a computer is the Coulomb gauge, which has the advantage of clearly isolating the physical transverse modes of the photon and allowing the longitudinal and gauge modes to be eliminated from the evolution problem, thereby making most efficient use of computer memory. In the Coulomb gauge the propagator and each of its pieces satisfies

$$\partial^m D_{m\nu} = \partial^m d_{m\nu} = \partial^m \widetilde{D}_{m\nu} = 0 \; . \qquad (4.24)$$

The Coulomb gauge condition substituted into (4.23) allows one to solve for the time components of the propagator $\widetilde{D}_{t\nu}$ explicitly in terms of the other components since there are no propagating timelike photons in this physical radiation gauge.

In cases of special symmetry, such as spatially homogeneous mean fields, considerable simplification of the equations of motion occur. In the scalar field case let $\chi = \chi(t)$ and $\phi = \phi(t)$ be functions only of time. Then the Wightman functions may be expressed as the Fourier transform,

$$G_>(t,\mathbf{x};t',\mathbf{x}') = -G_<^*(t,\mathbf{x};t',\mathbf{x}') = \int [d\mathbf{k}] e^{i\mathbf{k}\cdot(\mathbf{x}-\mathbf{x}')} G_>(t,t';\mathbf{k}) \qquad (4.25)$$

with
$$[d\mathbf{k}] \equiv \frac{d^3\mathbf{k}}{(2\pi)^3} \tag{4.26}$$

and $G_>$ a function only of $k \equiv |\mathbf{k}|$. Introducing the Fourier mode functions $f_k(t)$ satisfying
$$\left[\frac{d^2}{dt^2} + k^2 + \chi(t)\right] f_k(t) = 0 \tag{4.27}$$

and normalized by the Wronskian condition,
$$f_k \frac{df_k^*}{dt} - f_k^* \frac{df_k}{dt} = i \tag{4.28}$$

we may express the Wightman function in the form,
$$G_>(t,t';k) = if_k(t)f_k^*(t')(N(k)+1) + if_k^*(t)f_k(t')N(k) + 2i\mathcal{R}e\left(f_k(t)f_k(t')F(k)\right) \tag{4.29}$$

where $N(k)$ and $F(k)$ carry the interpretation of particle number and correlated pair density in the general spatially homogeneous initial state:
$$\begin{aligned}
\text{Tr}\left(\rho a_\mathbf{k}^\dagger a_{\mathbf{k}'}\right) &= (2\pi)^3 \delta^3(\mathbf{k}-\mathbf{k}')N(k) \\
\text{Tr}\left(\rho a_\mathbf{k} a_{\mathbf{k}'}\right) &= (2\pi)^3 \delta^3(\mathbf{k}+\mathbf{k}')F(k) \, .
\end{aligned} \tag{4.30}$$

By likewise Fourier transforming the $D$ propagator, the field equations for the spatially homogeneous case may be expressed in the final form,
$$\left(\frac{d^2}{dt^2} + \chi(t)\right)\phi(t) = \frac{1}{\pi^2 N} \int_0^t dt' \int_0^\infty k^2 dk \, \mathcal{I}m \, [G_>(t,t';k)D_>(t,t';k)]\phi(t') \, , \tag{4.31}$$

$$\chi(t) = \mu^2 + \frac{\lambda}{2}\phi^2(t) + \frac{\lambda}{4\pi^2} \int_0^\infty k^2 dk \, \mathcal{I}m \, G_>(t,t;k) + \tag{4.32}$$
$$\frac{2\lambda}{\pi^2 N} \int_0^t dt_1 \int_0^{t_1} dt_2 \int_0^\infty k^2 dk \, \mathcal{R}e \left[G_>(t,t_1;k)\right] \mathcal{I}m \left[\widetilde{\Sigma}_>(t_1,t_2;k)G_>(t_2,t;k)\right],$$

$$\begin{aligned}
\widetilde{D}_>(t,t';k) &= \lambda^2 \widetilde{\Pi}_>(t,t';k) + 2\lambda \int_0^{t'} dt_1 \widetilde{\Pi}_>(t,t_1;k) \mathcal{R}e \, \widetilde{D}_>(t_1,t';k) \\
&\quad -2\lambda \int_0^t dt_1 \left[\mathcal{R}e \, \widetilde{\Pi}_>(t,t_1;k)\right] \widetilde{D}_>(t_1,t';k)
\end{aligned} \tag{4.33}$$

with
$$\begin{aligned}
\widetilde{\Pi}_>(t,t';k) &= \phi(t)G_>(t,t';k)\phi(t') - \frac{i}{2}\int [d\mathbf{q}]G_>(t,t';q)G_>(t',t;|\mathbf{k}-\mathbf{q}|) \\
\widetilde{\Sigma}_>(t,t';k) &= -\phi(t)\widetilde{D}_>(t,t';k)\phi(t') + i\int [d\mathbf{q}]G_>(t,t';q)\widetilde{D}_>(t,t';|\mathbf{k}-\mathbf{q}|)
\end{aligned} \tag{4.34}$$

and eqs. (4.27)-(4.29) above. All the equations have been expressed in terms of the $G_>$ and $\widetilde{D}_>$ functions by using the relations (4.3) and (4.7).

The result of our derivation is a well-defined initial value problem posed by the closed set of causal equations of motion and initial data, (4.27)-(4.34), corresponding to the mean fields and their fluctuations evolving forward in time from a specified spatially homogeneous initial density matrix. These equations are now suitable for numerical solution, provided the ultraviolet divergences in the momentum integrations are absorbed into renormalized parameters of the theory in the usual way. It is to this technical issue of the renormalization procedure that we turn next.

# 5 Renormalization and Energy-Momentum Tensor

To lowest order in $1/N$ (which includes the time dependent Hartree-Fock or Gaussian approximation) a convenient approach to the identification and removal of ultraviolet divergences in the evolution equations is the adiabatic method. This method is based on the fact that the large momentum behavior of the mode functions satisfying the differential eq. (4.27) is determined by an asymptotic expansion in derivatives of the time dependent frequency $\sqrt{k^2 + \mathcal{X}(t)}$. The divergences in the currents appearing in the mean field equations for $\mathcal{X}(t)$ or $A^\mu(t)$ are contained in the first few terms of this adiabatic expansion for the mode functions, and coincide with the divergences of the manifestly covariant in-out formalism. Hence they are removed by the same counterterms as in the covariant approach.

At next order in $1/N$ the structure of the integro-differential equations for the mean fields is considerably more complicated, and the adiabatic method for identifying and explicitly removing the ultraviolet divergences appears to be quite unwieldy. In solving the evolution equations numerically, an ultraviolet (and infrared) cut-off is always present. There is no real need to remove the explicit cut-off dependence appearing in intermediate quantities, provided only that physical results are cut-off independent in the end. If all bare quantities are taken to depend on the explicit ultraviolet momentum cut-off and the renormalized parameters of the theory in the same way as in the usual covariant treatment, then the results for the physical time evolution of the system should be insensitive to the the cut-off in the final analysis. This insensitivity of the solution to the cut-off may be checked empirically by changing the cut-off and evolving the equations again with the same initial data. That is, we increase the cut-off and rescale the bare parameters keeping the renormalized parameters fixed until the evolution is insensitive to the cut-off. On the order of several thousand field modes are typically required to approach this regime of insensitivity to the cut-off.

To illustrate the practicality of this method consider the simplified case of spatially homogeneous mean fields in the lowest order of the $1/N$ expansion. The scalar $\Phi^4$ equations read in this case simply,

$$\left[\frac{d^2}{dt^2} + \mathcal{X}(t)\right]\phi(t) = 0,$$

$$\mathcal{X}(t) = \mu^2 + \frac{\lambda_\Lambda}{2}\phi^2(t) + \frac{\lambda_\Lambda}{4\pi^2}\mathcal{I}m\int_0^\Lambda k^2 dk\, G_>(t,t;k), \tag{5.1}$$

with $G_>(t,t;k)$ given in terms of the mode functions $f_k(t)$ by eqs. (4.27)-(4.29), and the dependence of the bare coupling upon the momentum cut-off is recorded explicitly. The $\mathcal{X}$ equation contains a quadratic divergence in the momentum integration as the cut-off $\Lambda \to \infty$, which is independent of time and must be compensated by a counterterm in the bare mass parameter $\mu^2$. This quadratic divergence may be removed by the simple device of expressing $\mathcal{X}(t)$ in terms of its finite initial condition at $t = 0$, i.e.,

$$\mathcal{X}(t) = \mathcal{X}(0) + \frac{\lambda_\Lambda}{2}\left(\phi^2(t) - \phi^2(0)\right) + \frac{\lambda_\Lambda}{4\pi^2}\int_0^\Lambda k^2 dk\, \mathcal{I}m\left\{G_>(t,t;k) - G_>(0,0;k)\right\}. \tag{5.2}$$

The integration over $k$ still leads to a logarithmic dependence on the cut-off for large $\Lambda$, but this is precisely compensated by the $\Lambda$ dependence of the bare coupling according to the usual renormalization framework, viz.,

$$\lambda_\Lambda = Z_\lambda^{-1}(\Lambda, m)\,\lambda_R(m^2) \tag{5.3}$$

with

$$Z_\lambda(\Lambda, m) = 1 - \frac{1}{16\pi^2}\lambda_R(m^2)\ln\left(\frac{\Lambda}{m}\right) = \left[1 + \frac{1}{16\pi^2}\lambda_\Lambda \ln\left(\frac{\Lambda}{m}\right)\right]^{-1} \tag{5.4}$$

and $\lambda_R(m^2)$ the renormalized quartic coupling defined at some finite mass scale $m^2$. Indeed, by dividing both sides of (5.2) by $\lambda_\Lambda$ and using eqs. (5.3) and (5.4), it is straightforward to verify that the logarithmic dependence on $\Lambda$ of the integral in (5.2) is cancelled by the logarithm in (5.4), so that the resulting equation for $\mathcal{X}(t)$ is independent of $\Lambda$ for $\Lambda$ large (provided $Z_\lambda > 0$). Since there is no $\phi$ (wavefunction) renormalization at lowest order in $1/N$ these simple steps are all that are required to arrive at cut-off independent evolution equations for the scalar theory at this order. Notice, in particular, that no adiabatic expansion of mode functions to isolate divergences is necessary in this approach. Such an expansion is useful only for verifying explicitly the cancellation of the cut-off dependence in (5.2) which must occur in any case.

In QED the Maxwell equation for the homogeneous mean electric field in the gauge $A_0 = 0$ and $A_i = \delta_{iz} A(t)$ at lowest order in $1/N$ becomes simply

$$\frac{d^2}{dt^2} A(t) = e_\Lambda^2 \operatorname{Tr}(\rho j^z(t)) = -\frac{ie_\Lambda^2}{2} \int [d\mathbf{k}] \operatorname{tr}\{[G_>(t,t;\mathbf{k}) + G_<(t,t;\mathbf{k})]\gamma^z\} \ . \qquad (5.5)$$

The Wightman functions $G_{>,<}$ may be expressed in terms of spinor mode functions obeying

$$\left[i\frac{d}{dt} - \alpha \cdot (\mathbf{k} - \mathbf{A}) - \beta m\right] \begin{pmatrix} u_{\mathbf{k}s}(t) \\ v_{\mathbf{k}s}(t) \end{pmatrix} = 0 \qquad (5.6)$$

through

$$\begin{aligned}
G_{>,<}(t,t';\mathbf{k}) &= \begin{cases} i\sum_s u_{\mathbf{k}s}(t)\overline{u}_{\mathbf{k}s}(t') \\ -i\sum_s v_{\mathbf{k}s}(t)\overline{v}_{\mathbf{k}s}(t') \end{cases} + \\
&\quad -i\sum_s N(\mathbf{k}s)\left(u_{\mathbf{k}s}(t)\overline{u}_{\mathbf{k}s}(t') - v_{\mathbf{k}s}(t)\overline{v}_{\mathbf{k}s}(t')\right) \\
&\quad +i\sum_s \left(F(\mathbf{k}s)u_{\mathbf{k}s}(t)\overline{v}_{\mathbf{k}s}(t') + F^*(\mathbf{k}s)v_{\mathbf{k}s}(t)\overline{u}_{\mathbf{k}s}(t')\right)
\end{aligned} \qquad (5.7)$$

respectively, where $N$ and $F$ are the mean number of particle pair and correlation densities in the initial state, and $s = 1,2$ labels the spin. By assumption, $N$ and $F$ fall off faster than $|\mathbf{k}|^{-4}$ for large $|\mathbf{k}|$, so that the only divergence on the right side of eq. (5.5) comes from the $N = F = 0$ (vacuum) contribution, viz.,

$$e_\Lambda^2 \langle 0|j^z(t)|0\rangle = \frac{e_\Lambda^2}{2} \int [d\mathbf{k}] \sum_s \{\overline{u}_{\mathbf{k}s}(t)\gamma^z u_{\mathbf{k}s}(t) - \overline{v}_{\mathbf{k}s}(t)\gamma^z v_{\mathbf{k}s}(t)\} \ . \qquad (5.8)$$

A naive cubic divergence in this expression is cancelled between the two charge conjugated spinors $u$ and $v$. There remains only the logarithmic cut-off dependence related to charge renormalization. Let

$$e_\Lambda^2 = Z_e^{-1}(\Lambda, m) e_R^2(m^2) \qquad (5.9)$$

with

$$Z_e(\Lambda, m) = 1 - \frac{1}{6\pi^2} e_R^2(m^2) \ln\left(\frac{\Lambda}{m}\right) = \left[1 + \frac{1}{6\pi^2} e_\Lambda^2 \ln\left(\frac{\Lambda}{m}\right)\right]^{-1} \qquad (5.10)$$

which is the usual covariant charge renormalization of QED to this order. By dividing both sides of the Maxwell equation of motion, (5.5) by $e_\Lambda^2$ and using the relations (5.9) and (5.10) it is straightforward to check that the logarithmic cut-off dependence of the current expectation value is precisely cancelled by the logarithmic $\Lambda$ dependence of (5.10). Numerical results demonstrating the cut-off independence of the final result for the time evolution are presented in Figs. 3. The adiabatic expansion of mode functions previously employed in Refs. [9][10][18] is therefore quite unnecessary.

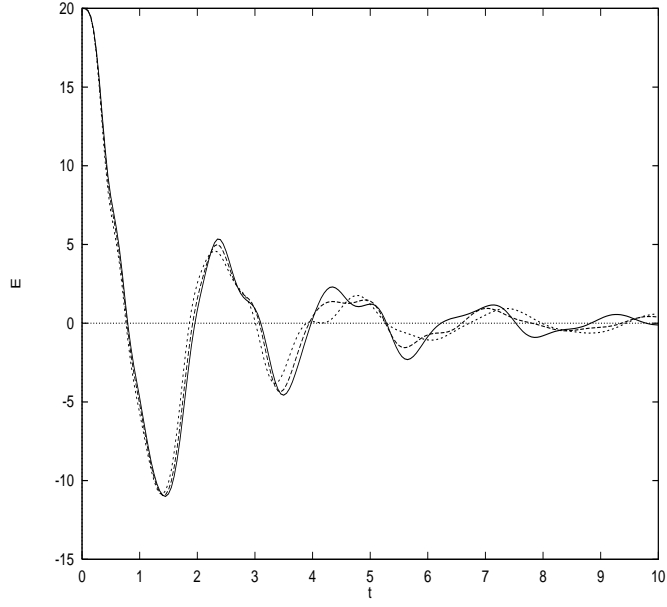

Figure 3: (a) Unrenormalized time evolution of the electric field in lowest order in $1/N$ for fixed initial conditions and fixed bare charge ($e^2 = 50$) but different values of the cutoff. The electric field is scaled by the critical field ($E \to eE/m^2$), and the time and cutoff are scaled by the mass (the cutoff is given in units of $m$ while time is measured in units of $m^{-1}$). The solid, dashed, and dotted lines correspond to cutoffs of 64, 48, and 32, respectively.

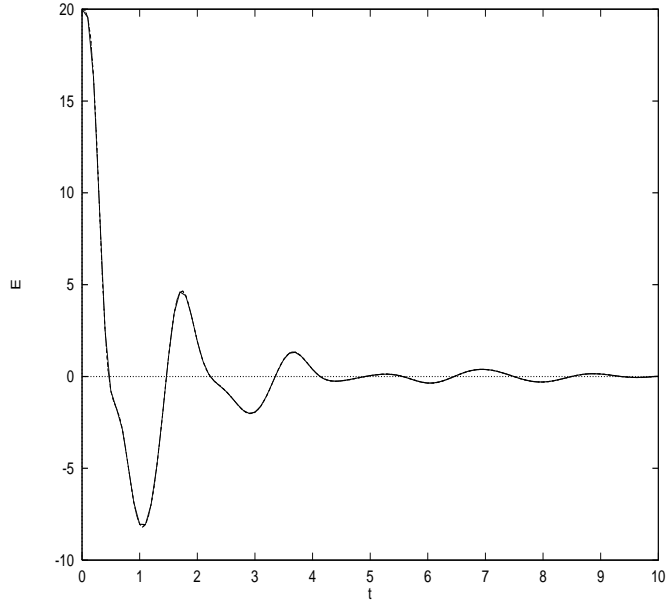

Figure 3: (b) Same evolution as (a) but with fixed renormalized charge ($e_R^2 = 50$), i.e., with the bare charge rescaled with cutoff according to equations (5.9) and (5.10). All three evolutions now fall on top of each other. Corresponding to the cutoff values of 64, 48, and 32, the values of $Z_e$ are .12, .18, and .26, respectively.

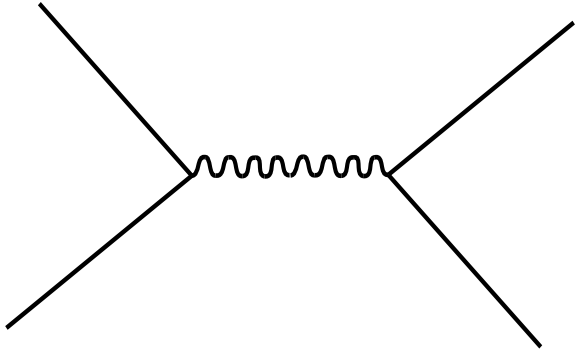

Figure 4: Born Diagram for $\mathcal{X}$ exchange which defines the renormalized quartic coupling $\lambda_R$ in the $1/N$ resummed $\Phi^4$ field theory.

The same strategy may be followed at any order of the $1/N$ expansion provided that the $\Lambda$ dependence of the bare couplings are known to the same order in the covariant treatment. This is an enormous technical simplification over standard adiabatic expansion methods for removing the cut-off dependence explicitly.

In $\lambda \Phi^4$ theory the general renormalization algorithm beyond lowest order requires knowledge of the $\phi^2 \mathcal{X}$ vertex and $\phi$-wavefunction renormalization constants $Z_1$ and $Z_2$ respectively, as well as the coupling renormalization $Z_\lambda$. The first two of these are defined in terms of the effective action by the conditions,

$$\begin{aligned} \Gamma(x,y;z) &\equiv -\frac{\delta^3 \mathcal{S}}{\delta\phi(x)\delta\phi(y)\delta\mathcal{X}(z)} = \frac{1}{Z_1}\Gamma_R(x,y;z) \\ \mathcal{G}^{-1}(x,y) &\equiv -\frac{\delta^2 \mathcal{S}}{\delta\phi(x)\delta\phi(y)} = \frac{1}{Z_2}\mathcal{G}_R^{-1}(x,y) \,, \end{aligned} \quad (5.11)$$

with $\Gamma_R$ the renormalized proper vertex part, and $\mathcal{G}_R^{-1}$ the renormalized inverse $\phi$ propagator in the case the mean field $\phi = 0$. In the case $\phi \neq 0$ we may continue to define $\mathcal{G}_R^{-1}$ as in (5.11) with the understanding that the two-point function for the fields $\phi$ and $\mathcal{X}$ becomes a *matrix* with off diagonal (*i.e.*, $\phi\mathcal{X}$) components, so that, in particular the renormalized $\phi$ propagator is no longer given simply by inverting $\mathcal{G}_R^{-1}$.

The other diagonal two-point vertex is

$$\mathcal{D}^{-1}(x,y) \equiv -\frac{\delta^2 \mathcal{S}}{\delta\mathcal{X}(x)\delta\mathcal{X}(y)} = \mathcal{D}_R^{-1}(x,y) \,, \quad (5.12)$$

which requires no renormalization, *i.e.*, it is already RG invariant, since it can be related directly to the physical Born scattering amplitude as discussed in the third paper of Ref. [16] and represented in Fig. 4. Indeed, the renormalized $\lambda$ coupling is specified by the Schwinger-Dyson equation for the $\mathcal{X}$ inverse propagator,

$$\lambda_R^{-1}(q^2) \equiv -\mathcal{D}^{-1}(q) = \frac{1}{\lambda_\Lambda} - \frac{i}{2}\int^\Lambda \frac{d^4p}{(2\pi)^4}\,\mathcal{G}(p+q)\,\Gamma(p,p+q)\,\mathcal{G}(p) \,, \quad (5.13)$$

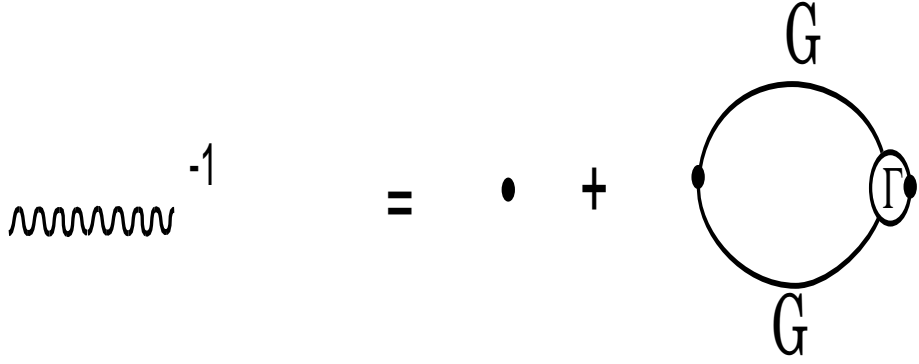

Figure 5: Exact Schwinger-Dyson equation for the $\chi$ inverse propagator $\mathcal{D}^{-1}$. Here $\mathcal{G}$ represents the exact unrenormalized $\phi$ field propagator and $\Gamma$ the exact unrenormalized vertex function. For the case of QED, the photon inverse propagator has a similar graphical representation where the first term is replaced by $d^{-1}$ and the vertex $\Gamma$ by $\Gamma_\mu$.

illustrated in Fig. 5. This defines the coupling renormalization $Z_\lambda(\Lambda, m = \sqrt{-q^2})$ in eq. (5.4) to any order of the $1/N$ expansion. To lowest order the vertex function, $\Gamma = \Gamma_0 = 1$ and the result of the previous discussion leading to eq. (5.4) is recovered.

Now, from the definitions (5.11) it follows immediately that

$$\begin{aligned}\Gamma(x,y;z) &= \frac{\delta}{\delta\chi(z)}\mathcal{G}^{-1}(x,y) = \frac{1}{Z_2}\frac{\delta}{\chi(z)}\mathcal{G}_R^{-1}(x,y)\\ &= \frac{1}{Z_2}\Gamma_R(x,y;z)\end{aligned} \quad (5.14)$$

since the renormalized vertex part $\Gamma_R$ may be defined by the $\chi$ variation of the renormalized self-energy appearing in $\mathcal{G}_R^{-1}(x,y)$. Comparing the last relation with the first member of eq. (5.11) implies the Ward Identity,

$$Z_1 = Z_2 \ . \quad (5.15)$$

Hence the vertex renormalization $Z_1$ may be computed from the self-energy renormalization $Z_2$ directly, just as in QED. Indeed, the definitions analogous to (5.11) in the QED case are

$$\begin{aligned}\Gamma^\mu(x,y;z) &\equiv -\frac{\delta^3 S}{\delta\psi(x)\delta\overline{\psi}(y)\delta A_\mu(z)} = \frac{1}{Z_1}\Gamma_R^\mu(x,y;z)\\ \mathcal{G}^{-1}(x,y) &\equiv -\frac{\delta^2 S}{\delta\psi(x)\delta\overline{\psi}(y)} = \frac{1}{Z_2}\mathcal{G}_R^{-1}(x,y) \ .\end{aligned} \quad (5.16)$$

The Ward Identity (5.15) follows immediately for the same reason as before. The general polarization tensor is guaranteed to be proportional to $q^2 g^{\mu\nu} - q^\mu q^\nu$ by the

Ward-Takahashi identity following from gauge invariance of the effective action, so that
$$q_\mu \mathcal{D}^{-1}(q)^{\mu\nu} = 0 , \tag{5.17}$$
and the renormalized charge may be defined by the Schwinger-Dyson equation for the photon inverse propagator,
$$\begin{aligned}\mathcal{D}^{-1}(q)^{\mu\nu} &\equiv \frac{1}{e_R^2(q^2)}(q^2 g^{\mu\nu} - q^\mu q^\nu) \\ &= d^{-1}(q)^{\mu\nu} - i\int^\Lambda \frac{d^4 p}{(2\pi)^4} \text{tr}\left\{\gamma^\mu \mathcal{G}(p)\Gamma^\nu(p,p+q)\mathcal{G}(p+q)\right\} ,\end{aligned} \tag{5.18}$$
illustrated again by Fig. 5, with the point vertex now representing $d^{-1\mu\nu}(q)$ and $\Gamma$ replaced by $\Gamma^\mu$. This defines the coupling renormalization $Z_e(\Lambda, m = \sqrt{-q^2})$ to any order of the $1/N$ expansion. To lowest order the vertex function, $\Gamma^\mu = \Gamma_0^\mu = \gamma^\mu$ and the result of the previous discussion leading to eq. (5.10) is recovered.

If there is no mean value of the Dirac field, the Ward Identity guarantees that wavefunction and vertex renormalization drops out of the Maxwell equations for the mean field and that the charge renormalization $Z_e(\Lambda, m)$ *alone* is sufficient to arrive at cut-off independent time evolution for the potential. To prove this we make use of several facts from the standard covariant analysis. First, we recall that the inverse propagator and vertex functions satisfy the Schwinger-Dyson equations,
$$\begin{aligned}\mathcal{G}^{-1}(q) &= G^{-1}(q) - \frac{i}{N}\int^\Lambda \frac{d^4 p}{(2\pi)^4}\gamma^\mu \mathcal{G}(p+q)\Gamma^\nu(p,p+q)\mathcal{D}_{\mu\nu}(p) \\ \Gamma^\mu(p,q) &= \gamma^\mu - \int \frac{d^4 r}{(2\pi)^4}\text{tr}\left\{\mathcal{G}(p+r)\Gamma^\mu(p+r,q+r)\mathcal{K}(p+r,q+r,p)\mathcal{G}(q+r)\right\} ,\end{aligned} \tag{5.19}$$
where $\mathcal{K}$ is the two particle irreducible scattering kernel. These equations are illustrated in Figs. 6. The diagrams contributing to $\mathcal{G}^{-1}$ and $\Gamma$ to first order in $1/N$ are illustrated in Figs. 7. The advantage of introducing the kernel $\mathcal{K}$ into the discussion is due to the fact that $\mathcal{GGK}$ is RG invariant, *viz.*,
$$\mathcal{G}\circ \mathcal{K}\circ \mathcal{G} = \mathcal{G}_R \circ \mathcal{K}_R \circ \mathcal{G}_R \tag{5.20}$$
in the condensed notation of Section 2. This means that we may eliminate the bare pointlike vertex $\gamma^\mu$ in favor of fully renormalized dressed quantities via
$$\gamma^\mu = \Gamma^\mu + \text{Tr}\left\{\Gamma^\mu \circ \mathcal{G}\circ \mathcal{K}\circ \mathcal{G}\right\} = \frac{1}{Z_1}\left(\Gamma_R^\mu + \text{Tr}\left\{\Gamma_R^\mu \circ \mathcal{G}_R \circ \mathcal{K}_R \circ \mathcal{G}_R\right\}\right) . \tag{5.21}$$
Then varying the exact Maxwell equation,
$$\partial_\nu F^{\mu\nu} = -ie^2 \text{Tr}\left\{\gamma^\mu \mathcal{G}[A]\right\} \tag{5.22}$$

$$G^{-1} = G^{-1} + \frac{1}{N} \;\text{(diagram)}$$

$$\Gamma = \cdot - \;\text{(diagram with } \Gamma \text{ and } K\text{)}$$

Figure 6: Exact Schwinger-Dyson equations for (a) $\mathcal{G}$ and (b) $\Gamma^\mu$ respectively.

$$G^{-1} = G^{-1} + \frac{1}{N} \;\text{(diagram)}$$

$$\Gamma = \cdot + \frac{1}{N}\left[\;\text{(diagram)} + \;\text{(diagram)}\;\right]$$

Figure 7: First two terms in the large $N$ expansion of (a) $\mathcal{G}$ and (b) $\Gamma^\mu$ respectively.

with respect to the mean potential and using the definitions (5.16) yields

$$\begin{aligned}
d^{-1\mu\nu}\delta A_\nu &= i\text{Tr}\left\{\gamma^\mu \mathcal{G} \circ \frac{\delta \mathcal{G}^{-1}}{\delta A_\nu} \circ \mathcal{G}\right\}\delta A_\nu \\
&= i\text{Tr}\left\{\gamma^\mu \mathcal{G} \circ \Gamma^\nu \circ \mathcal{G}\right\}\delta A_\nu \\
&= i\frac{Z_1^2}{Z_1^2}\text{Tr}\left\{(\Gamma_R^\mu + \text{Tr}\left\{\Gamma_R^\mu \circ \mathcal{G}_R \circ \mathcal{K}_R \circ \mathcal{G}_R\right\})\mathcal{G}_R \circ \Gamma_R^\nu \circ \mathcal{G}_R\right\}\delta A_\nu \\
&= i\text{Tr}\left\{(\Gamma_R^\mu + \text{Tr}\left\{\Gamma_R^\mu \circ \mathcal{G}_R \circ \mathcal{K}_R \circ \mathcal{G}_R\right\})\mathcal{G}_R \circ \Gamma_R^\nu \circ \mathcal{G}_R\right\}\delta A_\nu , \quad (5.23)
\end{aligned}$$

by eqs. (5.16), (5.15) and (5.21). The point of this exercise is that the Ward Identity guarantees the cancellation of wavefunction and vertex renormalization constants from the gauge invariant Maxwell equation for the mean potential. Since $Z_1/Z_2 = 1$ have cancelled out, the remaining loop integration over the skeleton diagrams of the last form of eq. (5.23) can have only log divergences which must be absorbed by charge renormalization alone. Indeed, comparing the second line of eq. (5.23) with the Schwinger-Dyson equation which defines the renormalized charge (5.18) shows that this is precisely what happens, since

$$\mathcal{D}^{-1\mu\nu}\delta A_\nu = \mathcal{D}_R^{-1\mu\nu}\delta A_\nu \qquad (5.24)$$

is the RG invariant linear response equation. Setting this quantity to zero allows us to study the dynamics of small perturbations in any given background mean field in a gauge covariant way independent of the cut-off. Since higher functional derivatives of the Maxwell equation involve no divergences whatsoever by simple power counting, we have proven that charge renormalization alone is sufficient to render the Maxwell equation fully RG invariant, and that the logic of the lowest order renormalization carries through to arbitrary order in $1/N$ without modification. One simply needs the corresponding expression for $Z_e$ to the given order of $1/N$, which has been given in the literature [20].

The analogous argument for the the $\lambda\Phi^4$ case encounters two complications. First, $\mathcal{X}$ contains quadratic divergences which must be handled correctly before applying the Ward Identity (5.15), and second, the mean field equation for $\phi$ will introduce the need for $\phi$ field renormalization over and above the coupling renormalization encountered in QED with zero mean Dirac fields. The first complication is handled by shifting from $\mathcal{X}$ to the renormalized mass, defined by

$$\mathcal{G}^{-1}(p^2 = -m_R^2) = 0 . \qquad (5.25)$$

To first order in $1/N$ (with $\phi = 0$) we have

$$m_R^2 = \mathcal{X} + \frac{1}{N}\Sigma(-m_R^2) = \mu^2 - \frac{i\lambda_\Lambda}{2}\int^\Lambda \frac{d^4p}{(2\pi)^4}\mathcal{G}(p^2) + \frac{1}{N}\Sigma(-m_R^2) . \qquad (5.26)$$

By definition $m_R^2$ is cut-off independent and RG invariant. Shifting to the physical mass pole is necessary in order to remove quadratic divergences on internal lines at order $1/N$ and higher. In the real time formalism the role of $m_R^2$ is played by the time dependent effective mass,

$$m^2(t) = \chi(t) - \frac{2}{N} \int_0^t dt' \int [d\mathbf{k}] \mathcal{I}m\{G(t,t';k)D(t,t';k)\} \qquad (5.27)$$

Since any potential quadratic divergence in $m^2(t)$ is cancelled by the time independent bare mass counterterm $\mu^2$, the quantity, $m^2(t) - m^2(0)$ is free of such divergences. At lowest order we were able to show this explicitly by eqs. (5.2)-(5.4). At next (and higher orders) the proof is a bit more involved since the quadratic divergences appear at the endpoint of the the time integration $t' \to t$ due to the singular nature of the integrand(s) in the coincident limit. Hence the divergent local counterterm as $t' \to t$ must be identified explicitly and removed before sensible results can be obtained. Since this counterterm is time independent it can be extracted most simply by adding and subtracting from the time dependent kernel, $\mathcal{I}m\{G(t,t';k)D(t,t';k)\}$ the same quantity evaluated at *constant* mass, which is the same as in the covariant treatment and may be calculated analytically. The difference $\mathcal{I}m\{G(t,t';k)D(t,t';k) - G_0(t,t';k)D_0(t,t';k)\}$ then has no quadratic divergence, while the remainder has a quadratic divergence structure which is known, and may be removed explicitly by integration by parts with respect to $t'$. The upper limit of this integration by parts (at $t' = t$) gives the local time independent quadratic cut-off dependence that is cancelled by subtracting $m^2(0)$, while the lower limit (at $t' = 0$) gives a finite cut-off independent term which oscillates rapidly (with the cut-off frequency) for small $t$. This high frequency "ringing" is a transient result of the "kick" to the system coming from our sharp initial condition at definite initial time which introduces high frequency components in the Fourier transform, and has been found in previous studies of quantum Brownian motion [4].

With these prior modifications to eliminate quadratic divergences explicitly, the argument leading to eq. (5.23) may be carried over to the scalar $\Phi^4$ case line for line, with the result that the evolution equation for $m^2(t) - m^2(0)$ is completely RG invariant, the remaining logarithmic divergences being absorbed by the $\lambda$ coupling renormalization alone with $Z_1/Z_2 = 1$ having cancelled from the expression, precisely as in the QED case.

The second difference of the $\Phi^4$ interaction from QED arises because of the existence of the non-vanishing $\phi$ mean field. Although the renormalization constants $Z_1$ and $Z_2$ drop out of the mass equation due to the Ward Identity (5.15), when the mean field $\phi$ is non-vanishing then it must be renormalized, so $Z_2 (= Z_1)$ will appear in the $\phi$ mean field equation. In the covariant formulation $Z_2$ is the logarithmic wave function renormalization constant given in terms of the derivative of the self-energy function on mass shell, namely

$$\frac{1}{Z_2} = \frac{\partial \mathcal{G}^{-1}(p^2)}{\partial p^2}\bigg|_{p^2=-m_R^2} = 1 + \frac{1}{N}\frac{\partial \Sigma(p^2)}{\partial p^2}\bigg|_{p^2=-m_R^2} + \mathcal{O}\left(\frac{1}{N^2}\right). \quad (5.28)$$

Since the field equation for the mean field involves $\mathcal{G}^{-1} \circ \phi$ and $Z_2^{-1}$ is a power series in $1/N$, the mean field equation is renormalized by multiplying only the first (order $1/N^0$) term of the $\phi$ eq. of motion (4.33) by $Z_2^{-1}$. That is,

$$\left(\frac{d^2}{dt^2} + m^2(t)\right)\phi(t)\frac{1}{Z_2} = \frac{1}{\pi^2 N}\int_0^t dt'\,(\phi(t) - \phi(t'))\int_0^\Lambda k^2 dk\,\mathcal{I}m\,[G_>(t,t';k)D_>(t,t';k)] \quad (5.29)$$

with $Z_2^{-1}$ given by the previous relation is the correctly renormalized equation of motion for the $\phi$ mean field to order $1/N$, with the $\Lambda$ dependence in $Z_2$ just cancelling that of the momentum integral in (5.29) to first order in $1/N$. Naturally, if non-zero mean charged fields are considered in electrodynamics they will have to be renormalized in the same way.

One additional remark about the $\lambda\Phi^4$ theory and its renormalization is in order. It is well known that this theory is trivial, in the sense that the cut-off cannot be removed to infinity without vanishing renormalized $\lambda$. From the point of view of practical numerical calculations this is irrelevant since a cut-off will always appear in the computer implementation, and the theory with finite cut-off is well defined and nontrivial. However, a necessary consequence of this point of view is that the cut-off cannot be removed in principle, and one should work in the range where the cut-off is not too large, or more precisely where

$$0 < 1 - Z_\lambda(\Lambda, m) = \frac{1}{16\pi^2}\lambda_R(m^2)\ln\left(\frac{\Lambda}{m}\right) + \mathcal{O}\left(\lambda_R(m^2)\ln\left(\frac{\Lambda}{m}\right)^2\right) < 1. \quad (5.30)$$

That this is the right condition on the cut-off may be seen either from triviality considerations or the senselessness of the theory when the Landau pole is reached at $Z_\lambda = 1$. Similar considerations presumably apply to QED as well, where the very weak coupling still affords an enormous range of momenta before the Landau pole is reached. So it is always possible to satisfy (5.30) and still have $\Lambda$ much greater than all frequencies of interest in the non-equilibrium time evolution of the fields, if the coupling is small enough. This necessary limitation on the theory has a positive side. If the quantity in (5.30) is small it is then permissible to develop the expression for $\tilde{D}$ in eq. (4.16) in a power series in $\lambda$, rather than solving this integral equation numerically, thereby recovering its ordinary perturbative expansion. This leads to an enormous economy of computer memory since the integral equation for $\tilde{\mathcal{D}}$ involves very big arrays and is extremely memory intensive. The evolution of the $\phi$ propagator and mean fields is still treated in the full $1/N$ expansion without modification. Further details of this procedure will be presented when we turn to numerical methods in future publications.

Finally, for many applications it is useful to have the energy-momentum tensor following from the quantum effective action by variation with respect to the metric of spacetime,

$$T_{\mu\nu} = -\frac{2}{\sqrt{-g}} \frac{\delta \mathcal{S}}{\delta g^{\mu\nu}} \,. \tag{5.31}$$

Since general coordinate invariance is maintained in the effective action, the energy-momentum tensor is conserved. This is an important property in nonequilibrium dynamics which is easily lost if one makes uncontrolled stochastic assumptions or approximations in a transport formulation. In the effective action approach, on the contrary, conservation of $T_{\mu\nu}$ is automatic, provided the cut-off procedure does not violate coordinate invariance.

In the scalar theory, after scaling out a factor of $N$ we have

$$\begin{aligned}\hat{T}_{\mu\nu} &= \partial_\mu \Phi \partial_\nu \Phi + g_{\mu\nu} \left\{ -\frac{1}{2} \partial^\alpha \Phi \partial_\nu \Phi - \frac{1}{2} \Phi^2 \chi + \frac{1}{\lambda} \chi \left( \frac{\chi}{2} - \mu^2 \right) \right\} \\ &+ \xi(g_{\mu\nu} \Box - \partial_\mu \partial_\nu) \Phi^2 \end{aligned} \tag{5.32}$$

in terms of the original quantum fields, where the last term proportional to the arbitrary parameter $\xi$ may be added without affecting conservation. By taking the expectation value of this quantity, we may express the conserved energy-momentum tensor in terms of the propagator and vertex functions introduced above, viz.,

$$\begin{aligned}T_{\mu\nu}(x) &= T^{cl}_{\mu\nu}[\phi,\chi](x) - i \left( \delta^\alpha_\mu \delta^\beta_\nu - \frac{1}{2} g^{\alpha\beta} g_{\mu\nu} \right) \partial_\alpha \partial'_\beta \mathcal{G}(x,x')|_{x=x'} \\ &- \frac{1}{2N} g_{\mu\nu} \int d^4x' \, d^4y' \, d^4z' \, \mathcal{D}(z',x) \mathcal{G}(x,x') \Gamma(x',y';z') \mathcal{G}(y',x) \\ &- \frac{i}{2\lambda N} g_{\mu\nu} \mathcal{D}(x,x) - i\xi \left( g_{\mu\nu} \Box - \partial_\mu \partial_\nu \right) \mathcal{G}(x,x) \,, \end{aligned} \tag{5.33}$$

where $T^{cl}_{\mu\nu}[\phi,\chi]$ is (5.32) evaluated on the mean fields, and it must be recalled that $\mathcal{G}$, the full connected two-point function for the $\Phi$ field is *not* the inverse of $\mathcal{G}^{-1}$ defined in eq. (5.11) when $\phi \neq 0$.

For QED the analogous expression is

$$\begin{aligned}T_{\mu\nu}(x) &= T^{cl}_{\mu\nu}[A] - \tfrac{i}{2} \nabla_{(\mu} \text{tr} \left\{ \gamma_{\nu)} \mathcal{G}(x,x) \right\} - \frac{i}{N} t^{\alpha\beta\rho\sigma}_{\mu\nu} \partial_\alpha \partial'_\beta \mathcal{D}_{\rho\sigma}(x,x')|_{x=x'} \\ &- \frac{1}{N} \int d^4x' \, d^4y' \, d^4z' \, \mathcal{D}_{\alpha(\mu}(z',x) \text{tr} \left\{ \gamma_{\nu)} \mathcal{G}(x,x') \Gamma^\alpha(x',y';z') \mathcal{G}(y',x) \right\} \end{aligned} \tag{5.34}$$

where

$$t_{\mu\nu}^{\alpha\beta\rho\sigma} \equiv \frac{1}{8} g_{\mu\nu}(g^{\alpha\sigma}g^{\beta\rho} - g^{\alpha\beta}g^{\rho\sigma}) + \frac{1}{4}(\delta^\rho_{(\mu}\delta^\sigma_{\nu)}g^{\alpha\beta} - \delta^\alpha_{(\mu}\delta^\beta_{\nu)}g^{\rho\sigma} + \delta^\rho_{(\mu}\delta^\beta_{\nu)}g^{\alpha\sigma} - \delta^\alpha_{(\mu}\delta^\sigma_{\nu)}g^{\beta\rho}) \quad (5.35)$$

and

$$T_{\mu\nu}^{cl}[A] = t_{\mu\nu}^{\alpha\beta\rho\sigma}\partial_\alpha A_\rho \partial_\beta A_\sigma = -\frac{1}{4}g_{\mu\nu}F^{\alpha\beta}F_{\alpha\beta} - F^\alpha_\mu F_{\alpha\nu} \quad (5.36)$$

is the stress tensor of the Maxwell mean potential. The next two terms on the first line of (5.34) are the contributions to the energy-momentum tensor of the fermions (moving in the mean potential $A$) and photons respectively, while the last term is the contribution of the interaction between them which appears first at order $1/N$.

Each energy-momentum tensor contains terms with quartic and quadratic dependence on the ultraviolet cut-off $\Lambda$. From general coordinate invariance of the effective action, these divergent contributions to $T_{\mu\nu}$ must be proportional to the metric of spacetime $g_{\mu\nu}$ which is flat here. Thus, these divergences may be isolated and removed rather easily by subtracting from the full $T_{\mu\nu}$ above the same quantity in zero mean field(s). The resulting subtracted $T_{\mu\nu}$ is still conserved and now completely finite. Indeed, the argument that the current appearing in the Maxwell equation (5.22) must be cut-off independent, provided that the bare charge is rescaled with the cut-off while keeping the renormalized charge fixed, may be taken over to the subtracted energy momentum tensor as well. Like the mass in the $\Phi^4$ theory, once the power law divergences are removed from $T_{\mu\nu}$ the resulting quantity is RG invariant, and may be interpreted as the physical energy-momentum of the nonequilibrium field theory evolution. Since this $T_{\mu\nu}$ is computed from the same propagator and vertex functions appearing in the evolution equations, the pressure, energy density and transport characteristics of the QED plasma or quantum $\Phi^4$ theory may be studied, and useful information about the approach to hydrodynamic behavior and/or an effective equation of state obtained. The effective equation of state for the QED plasma to lowest order in $1/N$ has been discussed in Ref. [10]. Detailed numerical results for these quantities and further applications of the large $N$ CTP method are in preparation and will be presented in future publications.

## 6 Conclusions

In this paper we have presented a general approach to the nonequilibrium evolution of a closed system of quantum fields. The large $N$ expansion permits a clean separation into mean fields and their fluctuations and constitutes a controlled approximation scheme to the infinite tower of coupled Schwinger-Dyson equations of quantum field theory. We have derived the equations from an effective action principle which preserves all classical symmetries (which are not anomalous) at the quantum level. This is important because of the central role invariances play in the renormalization procedure through the Ward Identities, which are in danger of being obscured in a noncovariant time evolution problem. Nevertheless, manifest causality of the time evolution is enforced by the Schwinger-Keldysh CTP formulation of the effective action principle. We have sketched a renormalization procedure involving an ultraviolet cut-off which is well-suited to numerical solution of the equations on a computer, and demonstrated explicitly the practicality of this scheme at lowest order of the $1/N$ expansion.

For definiteness, throughout the paper we have developed the general approach in the framework of two particular and familiar quantum field theories, *viz.*, $\lambda\Phi^4$ and QED. From this beginning there are three well-defined vectors for future work. The first consists in carrying out practical numerical computations in these theories for specific applications, for example, to the evolution of disoriented chiral condensates in heavy-ion collisions or to $e^+e^-$ particle production and shorting of strong fields in astrophysical plasmas. These applications are clearly interesting in their own right. The second direction to pursue is the use of these particular field theories as model systems for the study of more general phenomena, such as dissipation and decoherence in closed quantum systems. The emergence of an effective Boltzmann or transport equation description from fundamental time reversal invariant dynamics may be studied in these realistic field theories in a controlled way without additional stochastic assumptions. Finally, because of the existence of a gauge invariant action principle, the general method followed in this paper can be extended to non-abelian gauge theories such as QCD and gravitation without essential difficulty. This will make it possible to take into account consistently the backreaction of quantum fluctuations on the nonequilibrium evolution of a mean color or metric field and open up interesting applications in studies of the quark-gluon plasma, black hole decay, and cosmological models. We plan to take up each of these lines of research in subsequent publications.

# 7  Acknowledgments

We thank Alex Kovner for several helpful discussions. This work was supported by the U. S. Department of Energy at Los Alamos National Laboratory. S. H. was supported in part by the Air Force Office of Scientific Research. Y. K. wishes to thank the Center for Nonlinear Studies at Los Alamos National Laboratory for providing partial support. Numerical work was carried out on the CM-5 at the Advanced

Computing Laboratory at Los Alamos National Laboratory.